\documentclass[journal,transmag]{IEEEtran}

\usepackage[ansinew]{inputenc}
\usepackage{epsfig}
\usepackage{amsfonts}
\usepackage{amssymb}
\usepackage{amsmath,amsthm,mathtools,braket}
\usepackage{subfigure}
\usepackage[usenames]{color}
\usepackage{comment}
\usepackage{multirow}
\usepackage{psfrag}
\usepackage{soul}
\usepackage{booktabs}
\usepackage{algorithm,algorithmicx}
\usepackage{algpseudocode}

      \newcommand{\field}[1]{\mathcal{#1}}

      \renewcommand{\epsilon}{\varepsilon}
      \renewcommand{\phi}{\varphi}

      \newcommand{\ie}{{i.\,e.\ }}

      \let\originalleft\left
      \let\originalright\right
      \renewcommand{\left}{\mathopen{}\mathclose\bgroup\originalleft}
      \renewcommand{\right}{\aftergroup\egroup\originalright}
      \let\Right\right
      \let\Left\left
      \makeatletter
      \def\right#1{\Right#1\@ifnextchar#1{\!}{}}
      \def\left#1{\Left#1\@ifnextchar#1{\!}{}}
      \makeatother

      \DeclarePairedDelimiter\abs{\lvert}{\rvert}%
      \DeclarePairedDelimiter\norm{\lVert}{\rVert}%
        \makeatletter
        \let\oldabs\abs
        \def\abs{\@ifstar{\oldabs}{\oldabs*}}
        \let\oldnorm\norm
        \def\norm{\@ifstar{\oldnorm}{\oldnorm*}}
        \makeatother

        \theoremstyle{definition}
        
        \newtheorem{conjecture}{Conjecture}

        \newtheorem{problem}{Problem}
        
        \newtheorem*{rep@theorem}{\rep@title}
        \newcommand{\newreptheorem}[2]{%
          \newenvironment{rep#1}[1]{%
            \def\rep@title{#2 \ref*{##1}}%
            \begin{rep@theorem}}%
            {\end{rep@theorem}}}
        \newreptheorem{theorem}{Theorem}
        \newreptheorem{corollary}{Corollary}
        \theoremstyle{plain}

\begin{document}

\title{On the Interactions between Multiple \\ Overlapping WLANs using Channel Bonding}

\author{Boris Bellalta, Alessandro Checco, Alessandro Zocca, Jaume Barcelo\thanks{Copyright (c) 2013 IEEE. Personal use of this material is permitted. However, permission to use this material for any other purposes must be obtained from the IEEE by sending a request to pubs-permissions@ieee.org. B.~Bellalta and J.~Barcelo are with Universitat Pompeu Fabra, Barcelona, Spain; A.~Checco is with Hamilton Institute, Ireland; A.~Zocca is with Eindhoven University of Technology, The Netherlands. The research of the 1st and 4th authors was partially supported by the Spanish government (project TEC2012-32354), and by the Catalan Government (SGR2009\#00617). The research of the 2nd author was financially supported by the Science Foundation Ireland (grant 11/PI/11771). The research of the 3rd author was financially supported by The Netherlands Organization for Scientific Research (NWO) through the TOP-GO grant 613.001.012. Corresponding author: B. Bellalta. e-mail: boris.bellalta@upf.edu}}

\date{}

\maketitle

\markboth{IEEE Transactions on Vehicular Technology, 2015.}%
{Bellalta \MakeLowercase{\textit{et al.}}: On the Interactions between Multiple Overlapping WLANs using Channel Bonding}

\begin{abstract}
Next-generation WLANs will support the use of wider channels, which is known as channel bonding, to achieve higher throughput. However, because both the channel center frequency and the channel width are autonomously selected by each WLAN, the use of wider channels may also increase the competition with other WLANs operating in the same area for the available channel resources. In this paper, we analyse the interactions between a group of neighboring WLANs that use channel bonding and evaluate the impact of those interactions on the achievable throughput. A Continuous Time Markov Network (CTMN) model that is able to capture the coupled dynamics of a group of overlapping WLANs is introduced and validated. The results show that the use of channel bonding can provide significant performance gains even in scenarios with a high density of WLANs, though it may also cause unfair situations in which some WLANs receive most of the transmission opportunities while others starve.  
\end{abstract}


\IEEEpeerreviewmaketitle


\section{Introduction}

The number of multimedia devices, including smartphones, laptops and High Definition (HD) audio/video players, that access the Internet through deployed WLAN Access Points is increasing every day and everywhere. To improve the performance of WLANs, the use of wider channels ---compared to a single or basic $20$ MHz channel--- has been considered recently. This technique is commonly known as channel bonding~\cite{deek2011impact}.

The use of channel bonding in WLANs was introduced in the IEEE 802.11n amendment~\cite{IEEE80211n}, where two basic $20$ MHz channels can be aggregated to obtain a 40 MHz channel. The IEEE 802.11ac amendment~\cite{IEEE80211ac} further extends this feature by allowing the use of $80$ and $160$ MHz channels by grouping 4 and 8 basic channels, respectively. It is expected that future WLAN amendments, such as the IEEE 802.11ax, will continue to develop the use of wider channels~\cite{bellalta2015ieee}.

However, the use of channel bonding also increases the probability that WLANs operating in the same area will overlap (i.e., two WLANs overlap if they share at least one basic channel), which may cause severe performance degradation for some or all of them. This performance degradation is caused by the coupled dynamics that occur between the overlapping WLANs due to the \textit{listen-before-talk} characteristic of the CSMA/CA protocol. This effect may be particularly relevant in urban areas, where the high density of WLANs may impact the suitability of this approach. 

To better understand the coupled dynamics during the operation of overlapping WLANs using channel bonding and to evaluate their effects in terms of performance, we model the described scenario using a Continuous Time Markov Network (CTMN)~\cite{boorstyn1987throughput}. We show that the CTMN model is able to accurately capture the operation and the achievable throughput of each WLAN, despite considering a continuous backoff timer instead of the slotted backoff counter that is used in the IEEE 802.11 Distributed Coordination Function (DCF). Note that models of the DCF that assume that all nodes are able to listen all transmissions from other nodes, such as the model presented in~\cite{bianchi2000performance}, are not valid for the scenarios considered in this paper because this requirement does not hold in general.

The contributions of this paper are as follows: 
 
\begin{enumerate}
    \item We introduce a CTMN model that captures the coupled dynamics of multiple overlapping non-saturated WLANs. It allows to configure at each node the traffic load, the packet size, the backoff contention window (CW), the channel position and width, and the transmission rate.
    \item To improve the computational efficiency when solving the CTMN model, we reduce its number of states by aggregating the activity of all nodes that belong to the same WLAN. We refer to it as the WLAN-centric model.  
    \item We describe, model and categorise the interactions that occur between multiple overlapping WLANs, as well as capture their coupled operation using the WLAN-centric model. We also show that some of the interactions are similar to those that appear in single-channel CSMA/CA multi-hop networks.
    \item We formulate the optimal proportional fair channel allocation for WLANs when they use channel bonding, which gives us the upper bound performance for a group of overlapping WLANs in saturated conditions.  
    \item We evaluate numerically the performance achieved by a group of neighboring WLANs that use channel bonding as a function of the number of overlapping WLANs, the number of available basic channels and the set of channel widths when WLANs randomly choose both the channel center frequency and the channel width. We then compare the results with those obtained using the proposed optimal proportional fair channel allocation scheme. 
\end{enumerate}

The paper is structured as follows. First, we introduce some related work in Section~\ref{Sec:RelatedWorks}. In Section~\ref{Sec:SystemModel}, we describe the system model and all of the assumptions that are made. In Section~\ref{Sec:ThroughputModel}, we present and validate the analytical model. Section~\ref{Sec:WLANcentric} characterises the potential interactions between WLANs. It also describes the extension of the node-centric throughput model to a WLAN-centric model in order to improve the computational efficiency when solving it, and provide a more compact characterisation of the overall system as well. In Section~\ref{Sec:ChannelAllocation}, we introduce both the centralised and decentralised channel allocation schemes considered in this work. In particular, for the centralised case, we propose a waterfilling algorithm for allocating channels to a group of overlapping WLANs, as well as the hypothesis that result in the optimal proportional fair allocation. We present the results in Section~\ref{Sec:NumResults}, studying the effect that the quantity of available basic channels and of WLANs has on the system performance. Finally, the most important results of the paper are summarised in the conclusions, and several recommendations about the use of channel bonding in next-generation WLANs are provided.


\section{Related Work} \label{Sec:RelatedWorks}

Since most previous studies only focused on channel bonding, channel selection algorithms or continuous time CSMA/CA throughput models, we present the related work in three separate sections. To the best of our knowledge, only \cite{arslan2010auto} and \cite{herzen2013distributed} simultaneously consider the channel center frequency and channel width selection.

\subsection{Channel Bonding}

The performance gains and drawbacks of channel bonding in IEEE 802.11n WLANs are analysed experimentally in \cite{deek2011impact,arslan2010auto}, where the authors show that channel bonding results in: $i$) a lower SINR (Signal to Interference and Noise Ratio) due to the reduction of the transmission power per Hz each time the channel width is doubled, $ii)$ a lower coverage range because wider channels require higher sensitivity, $iii)$ a greater chance to suffer from and create interference, and $iv)$ more competition with other WLANs operating in the same area. However, they also show that channel bonding can provide significant throughput gains when those issues can be overcome by adjusting the transmission power and rate. 

The same considerations as in IEEE 802.11n are valid for channel bonding in IEEE 802.11ac. However, because it extends the channel bonding capabilities of IEEE 802.11n by allowing the use of $80$ and $160$ MHz channels, both the negative and positive aspects are accentuated. Therefore, there is much interest in developing effective solutions at both PHY and MAC layers to get the most benefit from channel bonding.  The performance of channel bonding in IEEE 802.11ac WLANs has been investigated by simulation in~\cite{Gong2011ChannelBounding, park2011ieee}, where both SBCA (Static Bandwidth Channel Access) and DBCA (Dynamic Bandwidth Channel Access) schemes are considered. The results presented in~\cite{Gong2011ChannelBounding, park2011ieee} show that channel bonding can provide significant throughput gains, but also corroborate the fact that these gains are severely compromised by the activity of the overlapping wireless networks. The impact of hidden nodes on the network performance in a specific scenario is evaluated in~\cite{Gong2011ChannelBounding}, where a protection mechanism based on the exchange of RTS/CTS frames is proposed. The sensitivity of the secondary basic channels and how the position of the primary basic channel affects the system performance are evaluated in~\cite{park2011ieee}. However, neither~\cite{Gong2011ChannelBounding} nor~\cite{park2011ieee} present any analytical model. Finally, channel bonding for short-range WLANs is considered in~\cite{bellaltachannel}, where the impact of other WLANs on the system performance is evaluated. 

\subsection{Channel Selection Algorithms}

Channel selection algorithms in wireless networks have been the subject of numerous investigations. The first studies on this topic focused on either centralised or distributed schemes that rely on message passing
(see for instance~\cite{leung2003frequency,mishra2006client,mishra2006distributed,raniwala2005architecture} and references therein). 

These schemes are not applicable in our case, however, because different WLANs generally have different administrative domains: indeed, as such, they are independent and autonomous systems. Channel bonding complicates the analysis even more because different groups of basic channels are used, which potentially makes communication between WLANs more difficult.

Several solutions, based mainly on graph theory, have been proposed trying to consider these constraints on communication. This solution requires decentralised algorithms for channel selection, see~\cite{clifford2007channel,kauffmann2007measurement,kauffmann2007self}.

Channel selection when wider channels are used has been considered only in~\cite{arslan2010auto} and~\cite{herzen2013distributed}. In~\cite{arslan2010auto}, the authors propose an algorithm to dynamically select the channel center frequency and to dynamically switch between a 20 or a 40 MHz channel width to maximise the throughput. However, the authors assume that the access points (APs) are able to exchange information (i.e., the achieved throughput on each channel) or that a central authority provides such information. A decentralised algorithm is proposed in~\cite{herzen2013distributed} to select both the channel center frequency and the channel width by sensing the interference that is caused by the other neighboring WLANs.

\subsection{Continuous Time CSMA/CA Models}  

The use of CTMN models for the analysis of CSMA/CA networks was originally developed in~\cite{boorstyn1987throughput} and was further extended in the context of IEEE 802.11 networks in~\cite{durvy2006packing,liew2010back,nardelli2012closed,laufercapacity,wang2005throughput}, among others. Although the modelling of the IEEE 802.11 backoff mechanism is less detailed than in Bianchi~\cite{bianchi2000performance}, it offers greater versatility in modelling a broad range of topologies. Moreover, the experimental results of~\cite{liew2010back,nardelli2012closed} demonstrate that CTMN models, even if stylized, provide remarkably accurate throughput estimates for actual IEEE 802.11 systems. A comprehensible example-based tutorial of CTMN models applied to different wireless networking scenarios can be found in~\cite{bellalta2014bookchapter}.

Boorstyn et al.~\cite{boorstyn1987throughput} introduce the use of CTMN models to analyse the throughput of multi-hop CSMA/CA networks and study several network topologies, including a simple chain, a star and a ring network. Wang et al.~\cite{wang2005throughput} extend the work in~\cite{boorstyn1987throughput} by considering also the fairness between the throughput achieved by each node, as well as providing several approximations with the goal of reducing the model complexity by using only local information. In addition, they relate the parameters of the CTMN model with those defined by the IEEE 802.11 standard, such as the contention window and the use of RTS/CTS frames. Durvy et al.~\cite{durvy2006packing} also use CTMN models to characterise the behaviour of wireless CSMA/CA networks and investigate their spatial reuse gain. Nardelly et al.~\cite{nardelli2012closed} extend previous models to specifically consider the negative effect of collisions and hidden terminals. They evaluate several multi-hop topologies and compare the results with experimental data to show that CTMN models can be very accurate. Liew et al.~\cite{liew2010back} validate the accuracy of CTMN to model CSMA networks using both simulations and experimental data. They also introduce a simple but accurate technique to compute the throughput of each node based on identifying the maximum independent sets of transmitting nodes. Recently, Laufer et al.~\cite{laufercapacity} extended such CTMN models to support non-saturated nodes and flow-based analysis of multi-hop networks. Finally, the CTMN model presented in~\cite{laufercapacity} is used in~\cite{bellaltaperformance} to evaluate the performance of a vehicular video surveillance system.


\section{System Model} \label{Sec:SystemModel}

\begin{figure}[t!]
\centering
\psfrag{a}[][][0.7]{$a$}
\psfrag{b}[][][0.7]{$b$}
\psfrag{c1}[][][0.7]{$c_1$}
\psfrag{c2}[][][0.7]{$c_2$}
\psfrag{d}[][][0.7]{$d$}
\psfrag{BSS1}[][][0.6]{WLAN A}
\psfrag{BSS2}[][][0.6]{WLAN B}
\psfrag{BSS3}[][][0.6]{WLAN C}
\psfrag{BSS4}[][][0.6]{WLAN D}
\psfrag{AP}[][][0.6]{AP}
\psfrag{STA}[][][0.6]{STA}
\epsfig{file=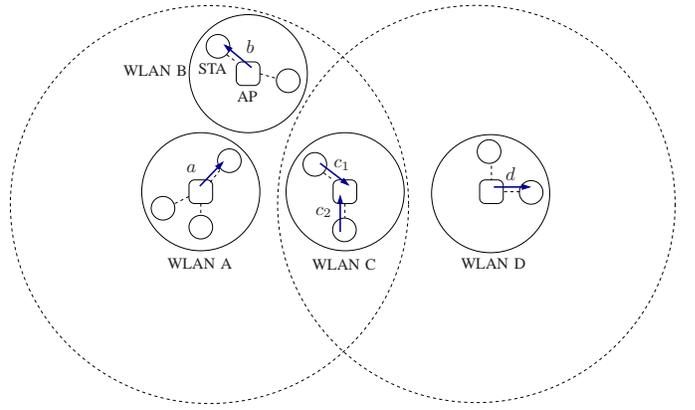,width=\columnwidth,angle=0}\label{Fig:RangesWLANs}\\
\caption{Two groups of neighboring WLANs (WLANs A, B and C in the one hand; and WLANs C and D in the other). The Data Communication Range (continuous line) and Carrier Sense Range (dashed line) are indicated in the plot. The two groups of neighboring WLANs interact because WLAN C belongs to both of them. Nodes $a$, $b$, $c_1$, $c_2$ and $d$ are transmitting a data flow.}
\label{Fig:RangesWLANsNoAdj_MH}
\end{figure}

In the following, we will say that a group of WLANs are neighbors when all of the WLANs are within the carrier sense range of the others. A WLAN may belong to several groups of neighboring WLANs, and therefore, those groups of neighboring WLANs may also interact between them through it (see Figure~\ref{Fig:RangesWLANsNoAdj_MH}). Table~\ref{Tbl:notation} summarizes the notation used in this paper.


\subsection{Network description}

We consider a system with $M$ WLANs spatially distributed over a certain area, where WLAN $i$ contains $U_i$ nodes, i.e., the AP and $U_i-1$ STAs. A set of $N$ predefined basic channels are at the disposal of all $M$ WLANs. When WLAN $i$ is initiated, or switches to a new channel, it selects a channel $C_i$ of width $W_i$, which is a contiguous subset of $c_i=|C_i|$ basic channels. If a basic channel has a width of 20 Mhz, then the width of channel $C_i$ is given by $W_i = 20 \cdot c_i$. The global set of channel allocations for the $M$ WLANs is $\mathcal{C}=\{C_1,\ldots,C_M\}$. We say that WLANs $i$ and $k$ \textit{overlap} if $C_i$ and $C_k$ share at least one basic channel, i.e., if $C_i \cap C_k\neq \emptyset$, given that both $i$ and $k$ are inside the carrier sense range of the other. In case two WLANs overlap, we assume they are outside the data communication range of the other, which makes the adjacent channel interference negligible. Finally, we also assume that the propagation delay between any pair of nodes is zero. 



\subsection{Node operation} 

\begin{figure}[t!]
\centering
\psfrag{T}[][][0.7]{T}
\psfrag{t}[][][0.7]{time}
\psfrag{Ch1}[][][0.6]{Ch. 1}
\psfrag{Ch2}[][][0.6]{Ch. 2}
\psfrag{Ch3}[][][0.6]{Ch. 3}
\psfrag{Ch4}[][][0.6]{Ch. 4}
\psfrag{Ch5}[][][0.6]{Ch. 5}
\psfrag{Ch6}[][][0.6]{Ch. 6}
\psfrag{Ch7}[][][0.6]{Ch. 7}
\psfrag{Ch8}[][][0.6]{Ch. 8}
\psfrag{ACK}[][][0.6]{ACK}
\psfrag{DATA}[][][0.7]{DATA}
\psfrag{AIFS}[][][0.7]{DIFS}
\psfrag{Backoff}[][][0.7]{Backoff}
\psfrag{External}[][][0.7]{Transmissions from other WLANs}
\epsfig{file=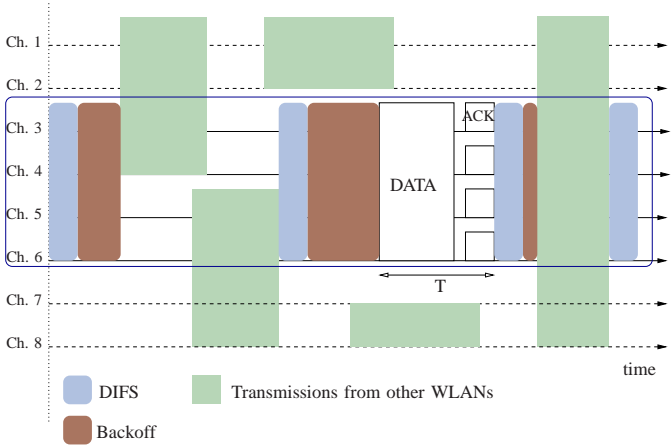,width=\columnwidth,angle=0}\label{Fig:ChannelBondingSimple}\\
\caption{Temporal evolution of the considered channel access scheme.}\label{Fig:Schemes2}
\end{figure}

The traffic load of node $j$ in WLAN $i$ is $\alpha_{i,j}$ packets/second. When a node has a packet ready for transmission, it checks the state of the channel $C_i$ that it has allocated. Once the channel has been sensed as being free for the duration of a DIFS (Distributed InterFrame Space), the node starts the backoff procedure by randomly initializing a timer. Every time a portion of the channel is detected as busy during the backoff interval, the backoff countdown is frozen until the entire channel width $W_i$ is detected as free again for the duration of a DIFS interval. This counter is decremented until it reaches zero, at which time the node starts transmitting a packet using the entire channel width $W_i$. Note that all nodes belonging to a group of neighboring WLANs will defer their backoff countdown accordingly if they share at least a basic channel with the transmitting node. Figure~\ref{Fig:Schemes2} shows the operation of the channel access for the specific case in which the target node uses four basic channels.

We assume that the backoff countdown at each node is in continuous time and has an average duration of $E[B_{i,j}]$ seconds for node $j$ in WLAN $i$. Therefore, when node $j$ has packets waiting for transmission, the attempt rate for every node is equal to $\lambda_{i,j}= E[B_{i,j}]^{-1}$.

The duration of a transmission of a packet by node $j$ in WLAN $i$ is denoted by $T_{i,j}(c_i,\gamma_{i,j},L_{i,j})$ and depends on the number $c_i$ of basic channels used, on the Signal-to-Noise Ratio (SNR) observed at the receiver side for that transmission, $\gamma_{i,j}$, and on the payload size $L_{i,j}$. Therefore, the packet departure rate, i.e., the rate at which packets depart from a node, is $\mu_{i,j}=E [T_{i,j}(c_i,\gamma_{i,j},L_{i,j})]^{-1}$. The probability that a packet is successfully received is $\eta_{i,j}$. We assume that the maximum number of retransmissions per packet is infinite. In this case, the effective number of packets per second that node $j$ has to transmit to successfully deliver its traffic load is $\alpha_{i,j}'=\frac{\alpha_{i,j}}{\eta_{i,j}}$. 


\begin{table*}[t!]
    \begin{tabular}{c|p{13cm}}
    \toprule
    \bf Notation & \bf Meaning \\
    \midrule
    $N$ & Number of available basic channels \\
    $M$ & Number of neighboring WLANs \\
    $U_i$ & Number of nodes in WLAN $i$ \\
    $C_i$ & Channel selected by WLAN $i$ \\
    $W_i$ & Width of the channel used by WLAN $i$ \\
    $\mathcal{C}=\{C_1,C_2,\ldots,C_M\}$ & Global channel allocation \\
    $c_i$ & Number of basic channels in $C_i$ \\    
    $\gamma_{i,j}$ & SNR observed at the receiver of node $j$ in WLAN $i$ transmissions \\
    $L_{i,j}$ & Size of the packets that node $j$ in WLAN $i$ transmits \\
    $T_{i,j}(c_i,\gamma_{i,j},L_{i,j})$ & Packet transmission duration from node $j$ in WLAN $i$ \\
    $\eta_{i,j}$ & Probability that a packet transmitted by node $j$ in WLAN $i$ is not received correctly\\    
    $\mu_{i,j}$ & Packet departure rate from node $j$ in WLAN $i$ \\
    $B_{i,j}$ & Duration of the backoff of node $j$ in WLAN $i$\\
    $\lambda_{i,j}$ & Backoff rate of node $j$ in WLAN $i$ given it has packets waiting for transmission \\
    $\theta_{i,j}$ & Activity ratio for node $j$ in WLAN $i$\\
    $\rho_{i,j}$ & Stationary probability that node $j$ in WLAN $i$ has packets to transmit when the channel $C_i$ is sensed idle\\    
    $\Omega(\mathcal{C})$ & Collection of all feasible network states \\
    $\pi_s$ & Steady-state probability of the network state $s \in \Omega(\mathcal{C})$ \\
    \bottomrule
    \end{tabular}
    \caption{Notation used in the system and analytical model.} \label{Tbl:notation}
\end{table*}

\subsection{Implications}

We discuss now the assumptions we have made on the node operation, and their implications for the results and conclusions in this work.

\begin{enumerate}
    \item \underline{No collisions with neighboring nodes}: Due to the choice of using a continuous-time backoff timer and to the fact that the propagation delay is assumed to be negligible, the probability of packet collisions between two or more nodes within the carrier sense range of the other nodes becomes zero. Therefore, the results we present could be considered as optimistic. However, for standard operating conditions and configurations, the collision probability in IEEE 802.11-based WLANs is also low, which makes this assumption very reasonable. The accuracy of such approximation has been extensively validated in previous works such as~\cite{liew2010back,laufercapacity}, and we will further validate it in Section~\ref{Sec:WLANcentric}. 
    
    Finally, it is worth to mention that this assumption allows us to easily model the interactions between nodes that are outside their carrier sense range because of the distance between them, or because they are operating in different channels. Other widely used IEEE 802.11-based WLANs analytical models such as those based on the works of Bianchi~\cite{bianchi2000performance} and Cali et al.~\cite{cali2000ieee} require that all nodes in the network are able to listen the transmissions from the others, and therefore they can not be applied in the scenarios considered in this work.
    \item \underline{No hidden nodes}: One key characteristic of IEEE 802.11 devices is that their carrier sense range is at least two times greater than their data range~\cite{deng2004tuning}. In this situation, the impact of hidden nodes is very low, as a given transmission can be only interfered by other transmissions from very distant nodes, with energy levels not higher than the noise floor. However, in specific deployments, where obstacles play also an important role on the propagation effects, hidden nodes may appear, and may severely affect the network performance~\cite{cai2013improving,jang2012ieee}. 
    \item \underline{Infinite Retransmissions}: In terms of the WLAN performance, allowing an infinite maximum number of retransmissions per packet does not affect much the final result because the probability that a packet is retransmitted more than few times is very low~\cite{chatzimisios2004performance}. However, such an assumption simplifies the analytical model as we do not need to keep track of the number of on-going retransmissions per packet.
\end{enumerate}


\section{Throughput Model}\label{Sec:ThroughputModel}

In this section, we introduce the Markovian model of the global system. In order to model the system as a Markov network, we assume that the durations of both the backoff and packet transmissions are exponentially distributed. Successively, we illustrate that, thanks to the insensitivity property of the Markov network, the results remain valid for more general probability distributions. Indeed, the insensitivity property guarantees that the throughput is insensitive to the distribution of the backoff and of the packet transmission duration, as it only depends on their expected value. 

\subsection{Continuous Time Markov Networks}

Suppose that a global channel allocation $\mathcal C = (C_1,\dots, C_M)$ for the $M$ WLANs is given. A \textit{feasible network state} is a subset of nodes that can transmit simultaneously, i.e., such that the WLANs to which they belong do not overlap. Let $\Omega(\mathcal{C})$ be the collection of all feasible network states. Note that any change in the global channel allocation $\mathcal{C}$ results in a different collection $\Omega(\mathcal{C})$ of feasible network states. 

    

Denote by $u_{i,j}$ node $j$ in WLAN $i$, with $j=1,\dots,U_i$. The local dynamics at every node described in previous section imply that the backoff rate of node $u_{i,j}$ is $\rho_{i,j}\lambda_{i,j}$, with $\rho_{i,j}$ the long-run stationary probability that node $j$ in WLAN $i$ has packets ready for transmission when the channel $C_i$ is sensed empty, and therefore the node is decreasing its backoff counter. The transmission rate of node $u_{i,j}$ is $\mu_{i,j} = 1 /E[T_{i,j}(c_{i},\gamma_{i,j},L_{i,j})]$. Then, the transition rates between two network states $s, s' \in \Omega(\mathcal{C})$ are
\begin{equation}
q(s,s')=
\begin{cases}
\rho_{i,j}\lambda_{i,j} & \text{ if } s'=s \cup \{u_{i,j}\} \in \Omega(\mathcal C),\\
\mu_{i,j} & \text{ if } s'=s \setminus \{u_{i,j}\},\\
0 & \text{ otherwise}.
\end{cases}
\end{equation}
{Denote by $S_t \in \Omega(\mathcal{C})$ the network state at time $t$. Thanks to the assumption on the backoff and transmission durations, $(S_t)_{t\geq 0}$ is a continuous-time Markov process on the state space $\Omega(\mathcal{C})$. This Markov process is aperiodic, irreducible and thus positive recurrent, since the state space $\Omega(\mathcal{C})$ is finite. Hence, it has a stationary distribution, which we denote by $\{\pi_s \}_{s \in \Omega(\mathcal C)}$.}

Let $\theta_{i,j}$ be the \textit{activity ratio} of node $u_{i,j}$, defined by
\[
\theta_{i,j}:=\frac{\rho_{i,j}\lambda_{i,j}}{\mu_{i,j}} = \frac{\rho_{i,j} E[T_{i,j}(c_{i},\gamma_{i,j},L_{i,j})]}{E[B_{i,j}]}.
\]
Note that $\theta_{i,j}$ depends on the number of basic channel $c_{i}$ assigned to WLAN $i$, since $\mu_{i,j}$ does. The process $(S_t)_{t\geq 0}$ has been proven to be a time-reversible Markov process in~\cite{kelly1979reversibility}. In particular, detailed balance applies and the stationary distribution $\{\pi_s \}_{s \in \Omega(\mathcal C)}$ of the process $(S_t)_{t \geq 0}$ can be expressed as a product form. The detailed balance relation for two adjacent network states, $s$ and $s \cup \{u_{i,j} \}$, reads
\begin{align}
\frac{\pi_{s\cup \{u_{i,j}\}}}{\pi_{s}} = \frac{\rho_{i,j}\lambda_{i,j}}{\mu_{i,j}}=\theta_{i,j}.
\end{align}
This relation implies that for any $s \in \Omega(\mathcal C)$
\begin{align}
\pi_{s} = \pi_{\emptyset} \cdot \prod_{u_{i,j} \in s} \theta_{i,j},
\end{align}
where $\emptyset$ denotes the network state where none of the nodes is transmitting. The last equality, together with the normalizing condition $\sum_{s \in \Omega(\mathcal{C})} \pi_s = 1$, yields
\begin{align}\label{Eq:product_form_empty}
\pi_{\emptyset} = \frac{1}{\sum_{s \in \Omega(\mathcal{C})} \prod_{u_{i,j} \in s} \theta_{i,j}} 
\end{align}
and
\begin{align}\label{Eq:product_form}
\pi_{s} = \frac{\prod_{u_{i,j} \in s} \theta_{i,j}}{\sum_{s \in \Omega(\mathcal{C})} \prod_{u_{i,j} \in s} \theta_{i,j}}, \quad s \in \Omega(\mathcal{C}).
\end{align}
Note that the normalizing constant $\pi_{\emptyset}$ and the stationary distribution $\{\pi_s \}_{s \in \Omega(\mathcal C)}$ depend on the state space $\Omega(\mathcal{C})$, and hence, they depend implicitly on the global channel allocation $\mathcal{C}$. 

Since the process $(S_t)_{t\geq 0}$ is irreducible and positive recurrent on $\Omega(\mathcal C)$, it follows from classical Markov chains results that $\pi_{s}$ is equal to the long-run fraction of time the system spends in the network state $s \in \Omega(\mathcal C)$.

\subsection{Packet Errors, Hidden Nodes \& External Interferers}

Packets can be received with errors. Errors are generally caused by the presence of ambient noise and interference. The sources of interference are diverse. We can define two main categories based on the use or not of the CSMA/CA rules by the interferer. If the interferer is operating under the CSMA/CA rules, we will refer to it either as a contender (i.e., the interferer is inside the carrier sense range of the transmitter) or as a hidden node (i.e., the interferer is outside the carrier sense range of the transmitter). Otherwise, we will simply classify it as an external interferer. 

The characterisation of the interference created by hidden nodes is complicated because of the coupled dynamics with the other nodes in the network, including also the one that suffers from the interference. In case of an external interferer, to characterise it we simply require its activity pattern. Assuming all those sources of errors are independent between them, we can define the probability that a packet transmitted by node $j$ in WLAN $i$ is successfully received as:
\begin{align}
    \eta_{i,j}=(1-p_{i,j}(\gamma_{i,j}))(1-p^h_{i,j})(1-p^{\text{ext}}_{i,j})
\end{align}
where $p_{i,j}(\gamma_{i,j})$ is the probability that a packet is corrupted due to ambient noise, $p^h_{i,j}$ is the probability that a packet is corrupted by a hidden node, and $p^\text{ext}_{i,j}$ the probability that it is corrupted by an external interferer.

There are several works that already consider the analysis of hidden nodes in Markov-based CSMA/CA network analysis, e.g., see~\cite{garetto2006modeling,nardelli2012closed} for further details. 

\subsection{Performance Metrics}

From the stationary distribution we compute the following performance metrics:

\begin{itemize}
    \item \textbf{Throughput}: the throughput $x_{i,j}(\mathcal{C})$ of node $j$ in WLAN $i$ for a given channel allocation $\mathcal{C}$ is
        \begin{align}\label{Eq:Throughput}
            x_{i,j} (\mathcal{C}) &:= \eta_{i,j} E[L_{i,j}] \mu_{i,j} \left(\sum_{s \in \Omega(\mathcal{C})\,:\, u_{i,j} \in s}{\pi_s}\right).
        \end{align}
    \item \textbf{Proportional Fairness}: The proportional fairness of the current channel allocation with respect to the throughput is 
        \begin{align}\label{Eq:Fairness}
            f(\mathcal{C}) :=\sum_{i=1}^{M}{\sum_{j=1}^{U_i}{\log{x_{i,j}(\mathcal{C})}}}.  
        \end{align}
    \item \textbf{Jain's Fairness index}: The Jain's Fairness Index (JFI) of the current channel allocation with respect to the throughput is 
        \begin{align}\label{Eq:Fairness}
            \field{J}(\mathcal{C}) :=\frac{\left(\sum_{i=1}^{M}{\sum_{j=1}^{U_i}{x_{i,j}(\mathcal{C})}}\right)^2}{\left(\sum_{i=1}^{M}{U_i}\right) \left(\sum_{i=1}^{M}{\sum_{j=1}^{U_i}{x_{i,j}^2(\mathcal{C})}}\right)}.  
        \end{align}        
\end{itemize}

\subsection{Computing the stationary distribution of the Markov network}


To compute the stationary distribution of the Markov network we need to compute all the $\rho_{i,j}$ values, i.e., $\{\pi_s \}_{s \in \Omega(\mathcal C)}=f(\boldsymbol{\rho})$, where $\boldsymbol{\rho}$ is a vector with all $\rho_{i,j}$ values respectively. However, in turn, their value depend also on the stationary distribution of the Markov network, i.e., $\boldsymbol{\rho}=g(\{\pi_s \}_{s \in \Omega(\mathcal C)})$. Thus, we have a set of non-linear equations, and in general, without a close-form solution.

To solve this set of non-linear equations, we have used an iterative fixed-point approach in which we update all the $\rho_{i,j}$ values until the throughput of all nodes converge to the solution. Note that if a node is not able to carry a load equal to its traffic load, i.e., $x_{i,j}(\mathcal{C})/E[L_{i,j}]=\alpha_{i,j}$, it will become saturated (i.e., $\rho_{i,j}=1$). 


\subsection{Solving the Model}

To solve the throughput model in a general scenario, we follow the next steps:

\begin{enumerate}
    \item We {fix} a global channel allocation $\mathcal{C}${, possibly generated at random.}
    \item {Starting from $\mathcal C$}, we compute all the overlaps $C_i \cap C_k$ between any two WLANs $i$ and $k$.
    \item We construct the collection $\Omega(\mathcal C)$ of all feasible network states. 
    \item We calculate the stationary probability $\pi_s$ for every network state $s \in \Omega (\mathcal C)$. 
    \item We calculate the throughput $x_{i,j}(\mathcal C)$ for every node $j$ in WLAN $i$ using the stationary distribution $\{\pi_s \}_{s \in \Omega(\mathcal C)}$.
    \item We compute the proportional fairness $f(\mathcal C)$ and Jain's Fairness index using the throughputs $x_{i,j}(\mathcal C)$, $i=1,\dots, M$ and $j=1,\dots, U_i$.
\end{enumerate}

\subsection{Numerical Example} \label{Sec:NumExampleModel}

\begin{figure}[t!]
\centering

\psfrag{Se}[][][0.9]{$s_{\emptyset}$}
\psfrag{sA}[][][0.9]{$s_{a}$}
\psfrag{sB}[][][0.9]{$s_{b}$}
\psfrag{sC1}[][][0.9]{$s_{c_1}$}
\psfrag{sC2}[][][0.9]{$s_{c_2}$}
\psfrag{sAC1}[][][0.9]{$s_{a,c_1}$}
\psfrag{sD}[][][0.9]{$s_{d}$}
\psfrag{sBD}[][][0.9]{$s_{b,d}$}

\psfrag{A}[][][0.8]{$a$}
\psfrag{B}[][][0.8]{$b$}
\psfrag{C1}[][][0.8]{$c_1$}
\psfrag{C2}[][][0.8]{$c_2$}
\psfrag{D1}[][][0.8]{$d$}
\psfrag{t}[][][0.8]{$t$}
\psfrag{Y(t)}[][][0.8]{$Y(t)$}
\epsfig{file=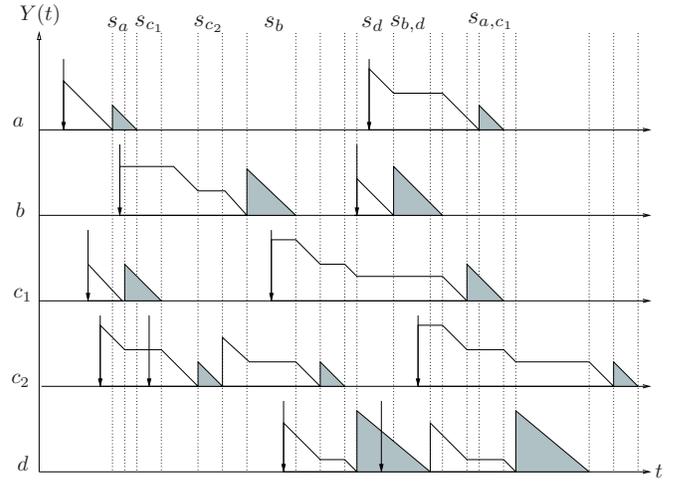,scale=0.325,angle=0}\\
\caption{Snapshot of the temporal evolution of the system considered in the example of Section \ref{Sec:NumExampleModel}. In the vertical axis, $Y(t)$ represents the amount of remaining backoff (white area) or transmission duration (blue area). The arrows inside the plot represent new packet arrivals.}\label{Fig:TempEvolCTMN}
\end{figure}

\begin{figure}[t!]
\centering
\psfrag{E}[][][0.9]{$\emptyset$}
\psfrag{A1}[][][0.9]{$a$}
\psfrag{B1}[][][0.9]{$b$}
\psfrag{C1}[][][0.9]{$c_1$}
\psfrag{C2}[][][0.9]{$c_2$}
\psfrag{D1}[][][0.9]{$d$}
\psfrag{A1C1}[][][0.9]{$a,c_1$}
\psfrag{A1C2}[][][0.9]{$a,c_2$}
\psfrag{A1D1}[][][0.9]{$a,d$}
\psfrag{B1D1}[][][0.9]{$b,d$}

\psfrag{a}[][][0.9]{$\{\rho_a\lambda_a,\mu_a\}$}
\psfrag{b}[][][0.9]{$\{\rho_b\lambda_b,\mu_b\}$}
\psfrag{c}[][][0.9]{$\{\rho_{c_1}\lambda_{c_1},\mu_{c_1}\}$}
\psfrag{d}[][][0.9]{$\{\rho_{c_2}\lambda_{c_2},\mu_{c_2}\}$}
\psfrag{e}[][][0.9]{$\{\rho_d\lambda_d,\mu_d\}$}
\psfrag{f}[][][0.9]{$\{\rho_{b}\lambda_{b},\mu_{b}\}$}
\psfrag{g}[][][0.9]{$\{\rho_{c_2}\lambda_{c_2},\mu_{c_2}\}$}
\psfrag{h}[][][0.9]{$\{\rho_{d}\lambda_{d},\mu_{d}\}$}
\psfrag{i}[][][0.9]{$\{\rho_{d}\lambda_{d},\mu_{d}\}$}
\psfrag{j}[][][0.9]{$\{\rho_{a}\lambda_{a},\mu_{a}\}$}
\psfrag{k}[][][0.9]{$\{\rho_{a}\lambda_{a},\mu_{a}\}$}
\psfrag{l}[][][0.9]{$\{\rho_{a}\lambda_{a},\mu_{a}\}$}
\psfrag{u}[][][0.9]{$\{\rho_{c_1}\lambda_{c_1},\mu_{c_1}\}$}
\psfrag{ll}[][][0.9]{$\{\rho_b\lambda_b,\mu_b\}$}
\epsfig{file=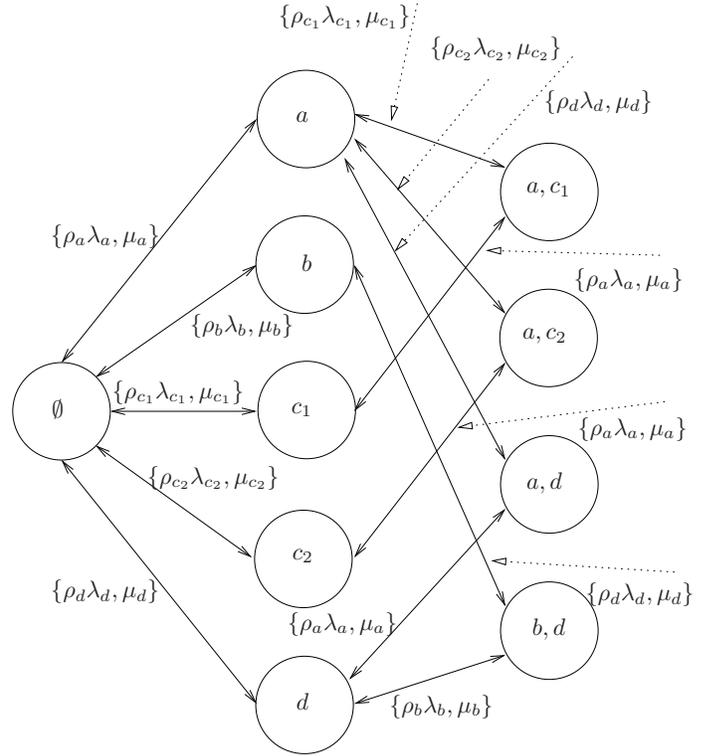,width=\columnwidth,angle=0}\\
\caption{CTMN model for the example of Section \ref{Sec:NumExampleModel}.}\label{Fig:ExCTMNCBonding}
\end{figure}

Let us consider the four neighboring WLANs shown in Figure~\ref{Fig:RangesWLANsNoAdj_MH}, and the following channel allocation for each one: $C_A=\{1,2,3,4\}$, $C_B=\{4,5\}$, $C_C=\{5,6,7,8\}$ and $C_D=\{5\}$. The network states in this scenario are $\Omega(\mathcal{C})=\{s_\emptyset,s_a,s_b,s_{c_1},s_{c_2},s_d,s_{a,c_1},s_{a,c_2},s_{a,d},s_{b,d}\}$, where $s_{\emptyset}$ is the network state in which none of the nodes is transmitting, $s_a$, $s_b$, $s_{c_1}$, $s_{c_2}$ and $s_d$ are the network states in which only node $a$, $b$, $c_1$, $c_2$ or $d$ is transmitting, respectively, and lastly $s_{a,c_1}$, $s_{a,c_2}$, $s_{a,d}$, and $s_{b,d}$ are the network states in which the two indicated nodes are simultaneously transmitting. Note that nodes $b$ and $d$ can transmit at the same time because they are outside the carrier sense area of the other though they have overlapping channels. Likewise, nodes $a$ and $c_1$ or $c_2$ can transmit simultaneously because they use non-overlapping channels in spite of being inside the carrier sense range of the other. A given snapshot of the temporal evolution of the five nodes is depicted in Figure~\ref{Fig:TempEvolCTMN}, where the different network states are separated by vertical dotted lines. The blue areas represent the time a node is transmitting and the white areas the time a node is in backoff. In case all nodes have exponentially distributed backoff and transmission times, this scenario can be modeled by the CTMN shown in Figure~\ref{Fig:ExCTMNCBonding}.

The stationary distribution from previous example is given by: $\pi_{a}=\theta_{a} \pi_{\emptyset}$, $\pi_{b}= \theta_{b} \pi_{\emptyset}$, $\pi_{c_1}= \theta_{c_1} \pi_{\emptyset}$, $\pi_{c_2}= \theta_{c_2} \pi_{\emptyset}$, $\pi_{d}= \theta_{d} \pi_{\emptyset}$, $\pi_{a,c_1}= \theta_{a}\theta_{c_1} \pi_{\emptyset}$, $\pi_{a,c_2}= \theta_{a}\theta_{c_2} \pi_{\emptyset}$, $\pi_{a,d}= \theta_{a}\theta_{d} \pi_{\emptyset}$, $\pi_{b,d}= \theta_{b}\theta_{d} \pi_{\emptyset}$, with $\pi_{\emptyset}=({1+\theta_{a}+\theta_{b}+\theta_{c_1}+\theta_{c_2}+\theta_{d}+\theta_{a}\theta_{c_1}+
\theta_{a}\theta_{c_2}+\theta_{a}\theta_{d}+\theta_{b}\theta_{d}})^{-1}$.

\begin{table*}[t!]
    \centering
    \begin{tabular}{lccccccc}
       \toprule                      
       & \multicolumn{4}{c}{\bf Parameters} & \bf Computation & \multicolumn{2}{c}{\bf Throughput per node [Mbps]}  \\ 
       \midrule                      
       & Node & $\alpha L$ [Mbps] & $E[T(c,\gamma,L)]$ [msecs] & $ p(\gamma) $ & $\rho$ & Analysis & Simulation \\
       \hline
       \multirow{5}{*}{Example 1} & $a$       &   18  &  0.1790   & 0.01 & 0.3673        &  18.00  & 17.97 \\
       & $b$       &   8   &  0.2070   & 0.10 & 0.3662        & 8.00 & 7.98 \\       
       & $c_1$   &   10  &  0.2150   & 0.05 & 0.6466        & 10.00 & 9.99 \\
       & $c_2$   &   22  &  0.1790   & 0.02 & 1.0000        & 15.95 & 15.75 \\
       & $d$       &   12  &  0.2630   & 0.15 & 0.6333        & 12.00 & 11.99  \\ 
       \hline   
       \multirow{5}{*}{Example 2} & $a$       &   4   &  0.1790   & 0.10 & 0.0744        & 4.00      & 3.9 \\
       & $b$       &   12  &  0.2070   & 0.10 & 0.3845        & 12.00        & 12.00 \\       
       & $c_1$   &   20  &  0.2150   & 0.15 & 1.0000        & 11.18       & 11.06 \\
       & $c_2$   &   5   &  0.1790   & 0.20  & 0.4752        & 5.00    & 5.00 \\
       & $d$       &   24  &  0.2630   & 0.05 & 1.0000        & 19.00       & 19.06 \\                           
       \bottomrule
    \end{tabular}
    \caption{Validation of the analysis in non-saturation conditions. A single simulation execution with a duration of 1000 seconds is considered for each example.} \label{Tbl:Example1}
\end{table*}

In order to validate the correctness of the presented analysis, we evaluate the described system considering the parameter values shown in Table~\ref{Tbl:Example1}, and compare the analysis results with simulations. In both cases, the backoff and the packet transmission duration are exponentially distributed. The rest of the considered parameters and their values, with the exception that in this example we are not considering packet aggregation ($N_{\text{a}}=1$), are shown in the Appendix~\ref{Sec:Parameters}, as well as information about the simulation tool used. The results are also shown in Table~\ref{Tbl:Example1}.

\subsection{Insensitivity}

For the Markov networks considered in this work, it turns out that the stationary distribution $\{\pi_s \}_{s \in \Omega(\mathcal C)}$ (and thus any analytic performance measure linked to it, such as the throughput) is insensitive to the distributions of backoff countdowns and transmission times, in the sense that it depends on these only through the ratios of their averages, i.e. $\theta_{i,j}$. The proof of the insensitivity result can be found in~\cite{liew2010back,van2010insensitivity}. The insensitivity property is crucial since back-off and transmission times may be not exponentially distributed in a real network.  
 

\section{Interactions between Overlapping WLANs} \label{Sec:WLANcentric}
  
The goal of this section is to characterise the different existing interactions between multiple overlapping WLANs. In order to do this, we first simplify the node-centric analytical model described in Section~\ref{Sec:ThroughputModel} by aggregating states. In this way, we reduce the total number of resulting network states and make its resolution more efficient. {Moreover, this new point of view} provides a more compact description of the interactions between neighboring WLANs. {Secondly,} we categorize the different types of interactions between WLANs, discussing how they impact on the performance of each one. We also show that some of those interactions are similar to those that appear in single-channel CSMA/CA multi-hop networks (see for example~\cite{boorstyn1987throughput,laufercapacity}). Lastly, we validate the use of the WLAN-centric analytical model by comparing its throughput predictions with the throughput values obtained from a detailed simulation of the same considered scenarios. Also, the presented numerical results give us some more insights about the interactions between WLANs. 
 
\subsection{WLAN-centric Throughput Analysis} 
 
Considering several WLANs with some active nodes in each results in a large number of network states, which requires large computation resources to solve the analytical model. Therefore, to make it more efficient, we simplify here the node-centric analytical model described in previous section by aggregating all those states in which the nodes of a given WLAN participate. We make also the following assumptions:

\begin{enumerate}
    \item We assume that all nodes in WLAN $i$ are close to each other and to the AP, and they observe similar SNR values. Therefore, they have a similar behavior from the point of view of a node belonging to another WLAN. 
    \item Considering non-saturation conditions, it is difficult to assess if the obtained results are due to the interaction between the different WLANs or the actual traffic load configuration of each node. To avoid such uncertainty, we assume from now on that all nodes are saturated, which can be considered as a worst-case scenario and will allow us to obtain more clear conclusions. Moreover, it also simplifies the development of the WLAN-centric model.
\end{enumerate}

Therefore, since the activity of each WLAN is the sum of the activity of its $U_i$ nodes, the WLAN-centric model is built based on the following considerations:

 
\begin{enumerate}
    \item A network state is now defined as the set of WLANs that are active simultaneously, instead of the set of nodes.
    \item The backoff rate of a WLAN $i$ is the sum of the backoff rates of all nodes in it, i.e., $\lambda_i=\sum_{j=1}^{U_i}{\lambda_{i,j}}$.
    \item The duration of a packet transmission in WLAN $i$ is $T_i(c_i,\gamma_{i},L_i)$, where $\gamma_i$ is the SNR observed by all packet transmissions inside WLAN $i$. Similarly, all nodes in WLAN $i$ transmit packets of size $L_i$ and have the same probability $\eta_i$ to receive a packet correctly.
    \item Since all nodes in WLAN $i$ are assumed to be saturated, WLAN $i$ is also saturated and $\rho_i=1$.
    \item The activity ratio of WLAN $i$ is given by $\theta_i=\lambda_i E[T_i(c_i,\gamma_{i},L_i)]$.
\end{enumerate}

To solve the model, the same approach as presented in Section~\ref{Sec:ThroughputModel} is considered. Also, because using the WLAN-centric model we can only compute the fraction of time a WLAN is active, the performance metrics previously described are modified accordingly. In Table~\ref{Tbl:NetVsNode} we compare both node and WLAN-centric models in terms of computational cost. Both models are executed in the same computer and using the same version of Matlab. The number of basic channels is set to $N=16$. Statistics are obtained by executing $200$ times each case. Each WLAN selects a channel width uniformly at random from the set of available channel widths $\{20,\ldots...,W_{\max}\}$ MHz. The position of the selected channel within the available channels is also picked uniformly at random. The throughput values converge by increasing the number of executions, which increases also the computation delay. It can be observed that both models give the same throughput but the node-centric one requires much more time and computational resources.

\begin{table*}
    \centering
    \begin{tabular}{lcccccccc}
    \toprule
     \multicolumn{4}{c}{\bf Parameters}  & \multicolumn{2}{c}{\bf States} & \multicolumn{2}{c}{\bf Agg. Throughput [Mbps]} & \bf Computation Delay [seconds] \\
     \midrule
    Model & M & U & $W_{\max}$ [MHz] & Mean & St. Dev. & Mean & St. Dev. & Mean \\
    \hline
    Node & 6 & 4 & 160 & 185.46 & 135.71 & 766 & 177 & 3.1\\
    WLAN & 6 & 4 & 160 & 30.53 & 9.85 & 774 & 168 &  0.4\\
    \hline 
    Node & 8 & 3 & 80 &  1195.4 & 855.9489 & 945 & 169 & 20.2 \\
    WLAN & 8 & 3 & 80 &  106.0 & 36.4014 & 966 & 177 & 1.6\\
    \hline
    Node & 12 & 2 & 40 &  20704.0 & 17967.0 & 1141.8 & 1646.2 & 395.4 \\
    WLAN & 12 & 2 & 40 &  738.7 & 356.3 & 1124.5 & 176.68 & 14.7\\
    \bottomrule    
    \end{tabular}
    \caption{Comparison between the number of states and the computation delay to obtain the stationary distribution between the node-centric and WLAN-centric approaches.}\label{Tbl:NetVsNode}
\end{table*}

\subsection{Cases of interest}

To illustrate the different cases of interest, we consider three neighboring WLANs, $A$, $B$ and $C$. All three WLANs transmit packets of fixed size ($L$), have the same number of nodes ($U$), have backoffs with the same average duration ($E[B]=\lambda^{-1}$), and use the same modulation and coding rate regardless of the number of basic channels selected by each WLAN. Therefore, if two WLANs use the same number of basic channels, the duration of a transmission is the same in both cases. Thus, for clarity, in the notation of time durations and activity ratios in this subsection, we will drop the subscript $i$ (which distinguishes the WLANs) and instead explicitly write the number of basic channels $c_i$ assigned to WLAN $i$.

\subsubsection{To overlap or not to overlap}

In the first example, we show that in terms of the throughput, the best option for all neighboring WLANs is to use non-overlapping channels. To illustrate this, we first consider the case in which all three WLANs use the same basic channels, namely $C_A=C_B=C_C=\{1,2,3,4,5,6\}$. Therefore, the set of feasible network states is $\Omega(\mathcal{C})=\{\emptyset,s_A,s_B,s_C\}$. The throughput achieved by WLAN $A$ is
\begin{align}
    x_A = \frac{L}{E[T(6)]} \pi_{s_A} =  \frac{\frac{L}{E[T(6)]} \theta(6)}{1+3 \theta(6)} = \frac{U \lambda L}{1+3 \theta(6)}.\nonumber 
\end{align}

By symmetry, the throughput achieved by each WLAN is identical and therefore

\begin{align}
    x_A = x_B=x_C = \frac{U \lambda L}{1+3 \theta(6)}.\nonumber 
\end{align}

Now consider a different scenario in which each WLAN uses two non-overlapping channels, namely $C_A=\{1,2\}$, $C_B=\{3,4\}$, and $C_C=\{5,6\}$. For this new channel allocation, the set of feasible network states is $\Omega(\mathcal{C})=\{\emptyset, s_A,s_B,s_C,s_{AB},s_{AC},s_{BC},s_{ABC}\}$. In this case, each WLAN is completely independent of the others and the network can therefore be modelled as three different systems. The throughputs achieved by the WLANs are again equal and are given by
\begin{align}
    x'_A = x'_B = x'_C = \frac{U \lambda L}{1+\theta(2)}.\nonumber 
\end{align}

Therefore, using WLAN $A$ as a reference, we can study the cases in which the achieved throughput when all WLANs overlap is better than the case in which each WLAN uses a non-overlapping set of channels. Because $x_A$ and $x_A'$ have the same numerator in both cases, the case in which all WLANs overlap will be better if $1+3\theta(6) < 1+\theta(2)$, or, equivalently $T(6) < T(2)/3$. Due to the channel access protocol defined in Section \ref{Sec:SystemModel}, the latter inequality will never hold, because the duration of some headers and other protocol overheads is not affected by the channel width.  

\subsubsection{Performance Anomaly}

The performance anomaly in multi-rate WLANs is well known \cite{heusse2003performance}. Due to the channel access mechanism, which is fair in terms of transmission opportunities, all nodes are able to transmit the same number of packets on average per unit of time, and therefore the nodes that are able to transmit at a fast rate are severely affected by nodes that can only transmit at a low rate. A similar result is observed when several WLANs overlap if they are using different number of basic channels. 

Consider three overlapping WLANs, $A$, $B$ and $C$, with the following channel allocations: $C_A=\{1,2,3,4\}$, $C_B=\{4,5\}$ and $C_C=\{4\}$. Despite the different channel widths, all three WLANs achieve the same throughput, which is given by:
\begin{align}
    x_A = x_B = x_C = \frac{U \lambda L}{1+\theta(4)+\theta(2)+\theta(1)} \nonumber 
\end{align}
which confirms the performance anomaly that was previously described.

The performance anomaly can be solved in several ways. For instance, the WLANs that use a wider channel can be allowed to transmit larger packets, so the overall transmission duration in all WLANs is the same. Alternatively, a different backoff duration can be assigned to each WLAN to guarantee that the WLANs that use more basic channels transmit more often.

\subsubsection{Non-direct interactions}

In this last example, we consider the case in which the performance of two WLANs that do not overlap is affected by the presence of a third WLAN. Suppose again that there are three WLANs, $A$, $B$ and $C$, and that the channel allocation is $C_A=\{1,2,3,4\}$, $C_B=\{5,6,7,8\}$ and $C_C=\{4,5\}$. In this scenario, the set of feasible network states is $\Omega(\mathcal{C})=\{\emptyset, s_A,s_B,s_C,s_{AB}\}$. The throughput achieved by WLAN $A$ is
\begin{align}
    x_{A}&=\frac{L}{E[T(4)]}(\pi_{s_A} + \pi_{s_{AB}}) =  \frac{ \frac{L}{E[T(4)]} \left( \theta(4)+\theta(4)^2 \right) }{1+\theta(2)+2\theta(4)+\theta(4)^2} \nonumber \\ &= \frac{ U \lambda L \cdot(1+\theta(4))}{1+\theta(2)+2\theta(4)+\theta(4)^2}, \nonumber
\end{align}
and, because $x_B=x_A$ by symmetry, the throughput achieved by WLANs B and C is
\begin{align}
    x_{B}&=x_{A} =\frac{ U \lambda L \cdot (1+\theta(4))}{1+\theta(2)+2\theta(4)+\theta(4)^2} \nonumber \\
    x_{C}&=\frac{ U \lambda L }{1+\theta(2)+2\theta(4)+\theta(4)^2}. \nonumber    
\end{align}

WLAN $A$ benefits from the existence of WLAN $B$, and vice versa, because they implicitly cooperate to starve WLAN $C$ in the competition for the channel resources. WLAN $C$ can only transmit when WLANs $A$ and $B$ are both silent. 

\subsection{Numerical Example}

\begin{figure}[ttttttt!]
\psfrag{BSS1}[][][0.6]{WLAN A}
\psfrag{BSS2}[][][0.6]{WLAN B}
\psfrag{BSS3}[][][0.6]{WLAN C}
\psfrag{BSS4}[][][0.6]{WLAN D}
\centering
\epsfig{file=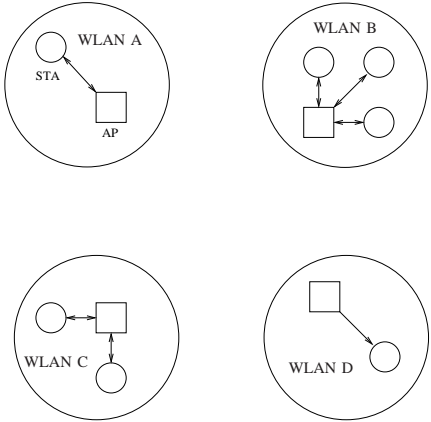,scale=0.4,angle=0}\\
\caption{A group of four neighboring WLANs. Arrows represent active traffic flows.}\label{Fig:ToyNetworkBasic}
\end{figure}

\begin{figure}[ttth!]
\psfrag{f}[][][0.7]{f}
\psfrag{WLAN A}[][][0.65]{WLAN A}
\psfrag{WLAN B}[][][0.65]{WLAN B}
\psfrag{WLAN C}[][][0.65]{WLAN C}
\psfrag{WLAN D}[][][0.65]{WLAN D}

\psfrag{1}[][][0.65]{1}
\psfrag{2}[][][0.65]{2}
\psfrag{3}[][][0.65]{3}
\psfrag{4}[][][0.65]{4}
\psfrag{5}[][][0.65]{5}
\psfrag{6}[][][0.65]{6}
\psfrag{7}[][][0.65]{7}
\psfrag{8}[][][0.65]{8}
\psfrag{9}[][][0.65]{9}
\psfrag{10}[][][0.65]{10}
\psfrag{S1}[][][0.85]{Scenario 1}
\psfrag{S2}[][][0.85]{Scenario 2}
\psfrag{S3}[][][0.85]{Scenario 3}
\psfrag{S4}[][][0.85]{Scenario 4}
\centering
\epsfig{file=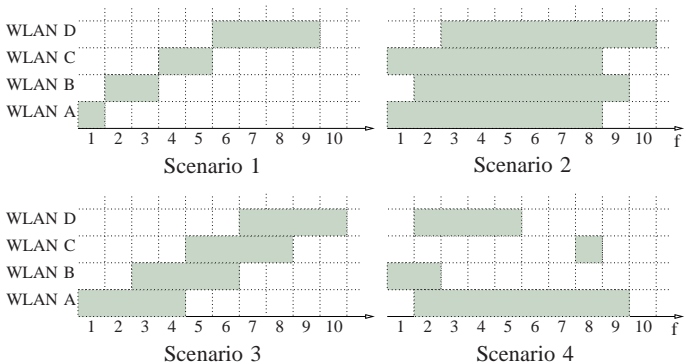,width=\columnwidth,angle=0}
\caption{Channel allocations.}\label{Fig:ChannelAssignment}
\end{figure}

Let us consider the network that is composed of four neighboring WLANs shown in Figure~\ref{Fig:ToyNetworkBasic}, and the four different channel allocations shown in Figure~\ref{Fig:ChannelAssignment}, which represent the non-overlapping (Scenario 1), fully overlapping (Scenario 2), WLAN in the middle (Scenario 3) and random channel selection (Scenario 4) scenarios, respectively. The number of available basic channels is set to $N=10$. 

\begin{figure}[ttth!]
\centering
\subfigure[Scenario 1]{\epsfig{file=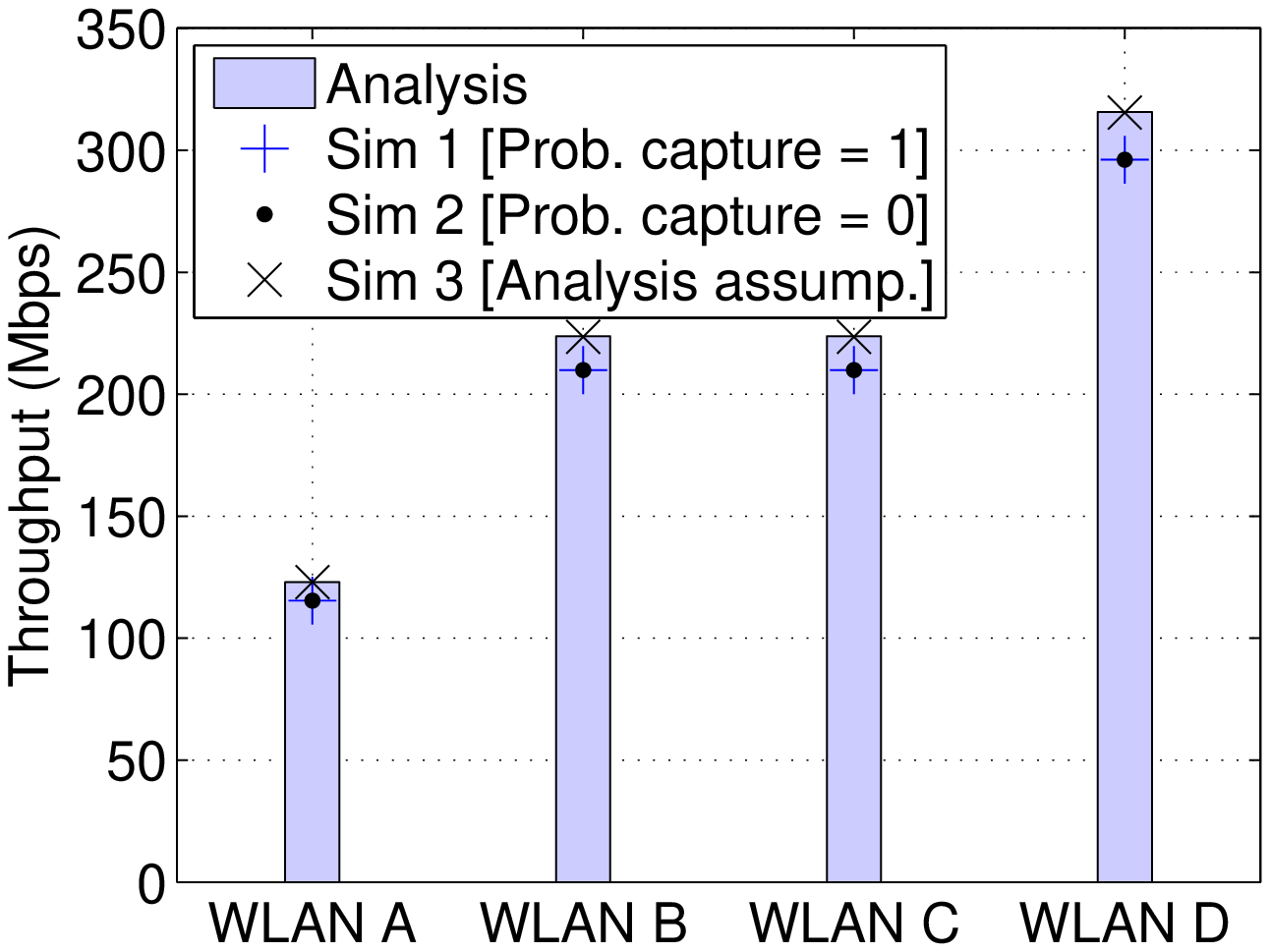,scale=0.32,angle=0}}\hspace{-0.5cm} 
\subfigure[Scenario 2]{\epsfig{file=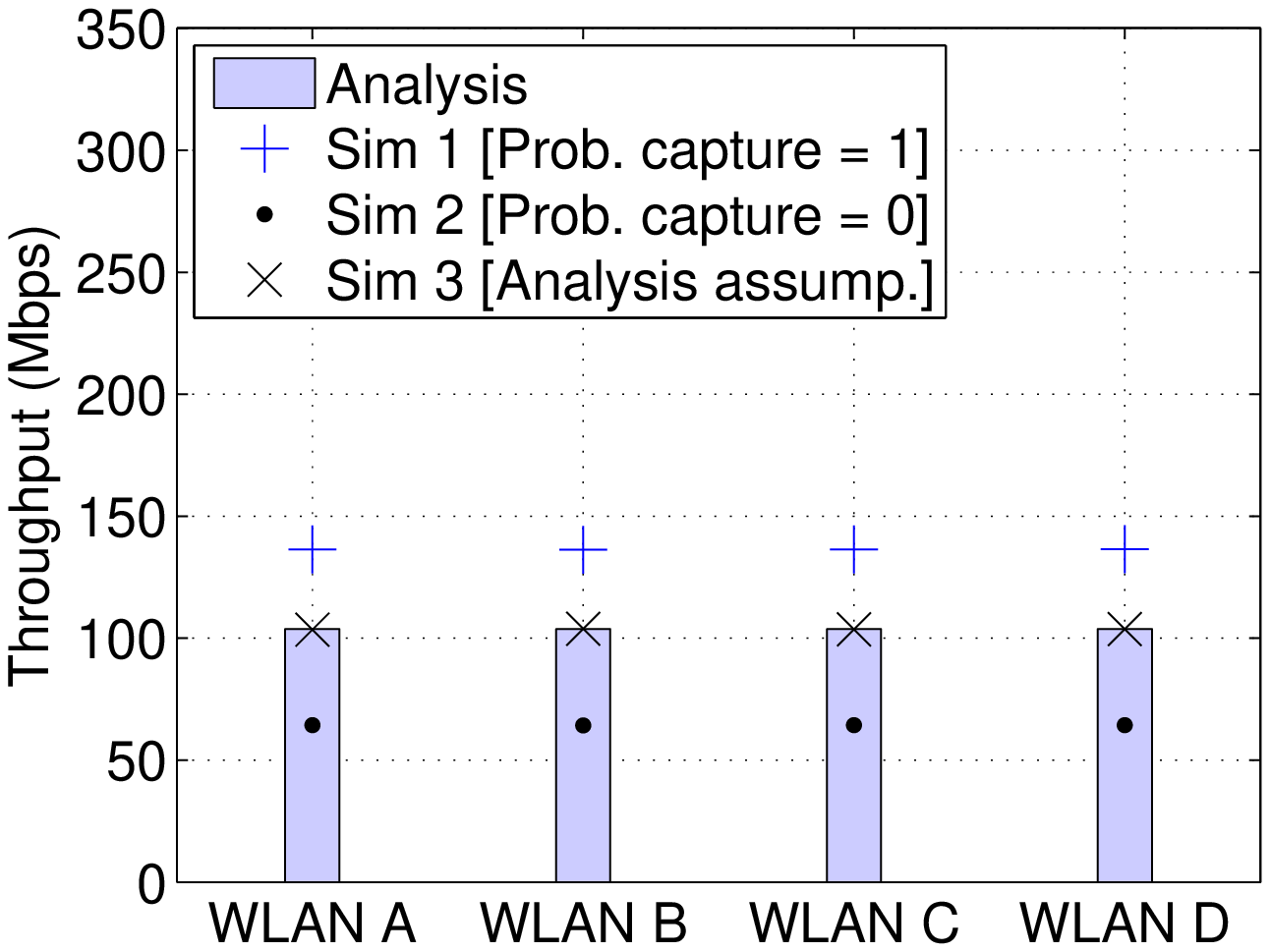,scale=0.32,angle=0}}\\ \vspace{-0.25cm}
\subfigure[Scenario 3]{\epsfig{file=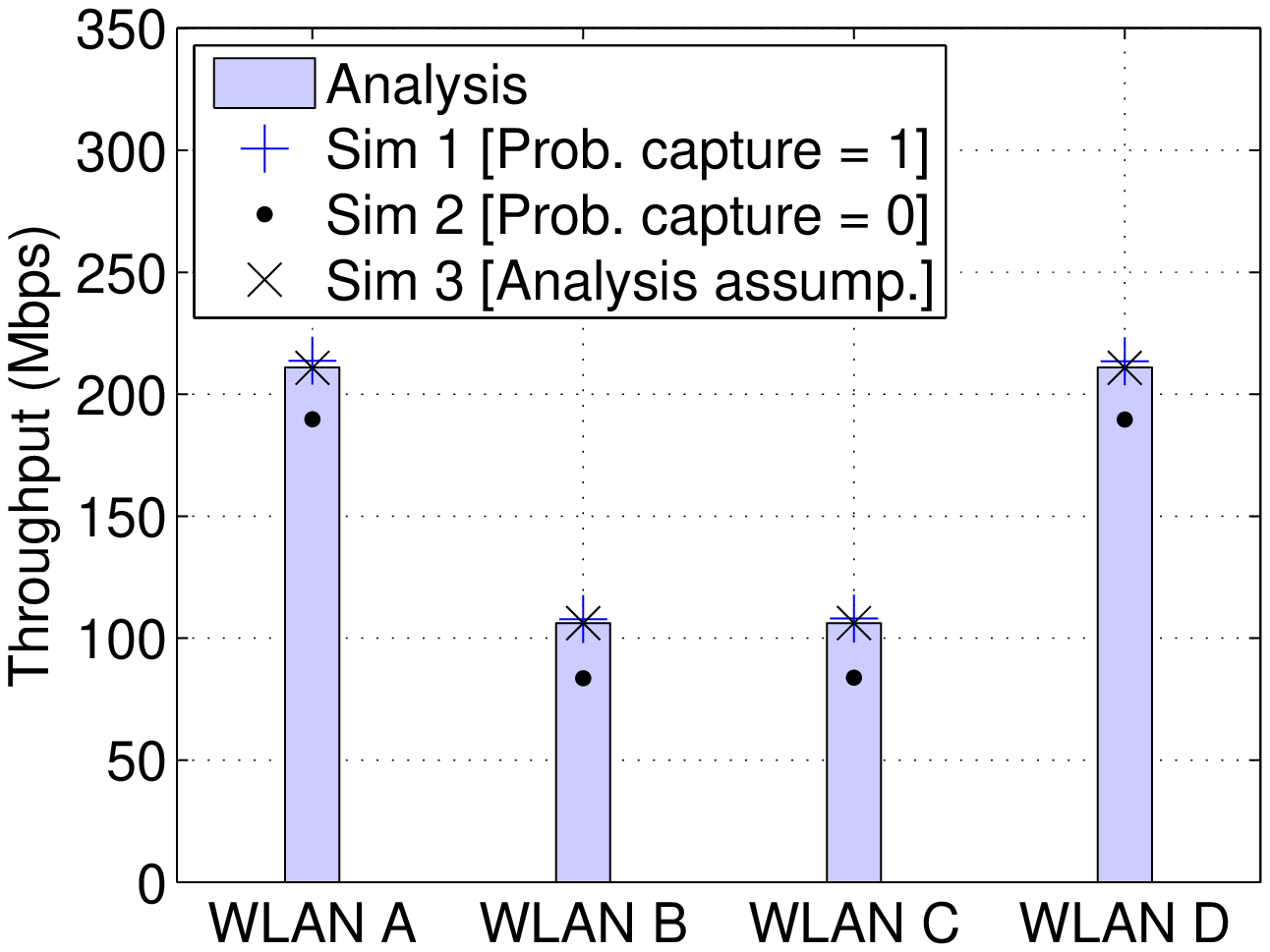,scale=0.32,angle=0}}\hspace{-0.5cm}
\subfigure[Scenario 4]{\epsfig{file=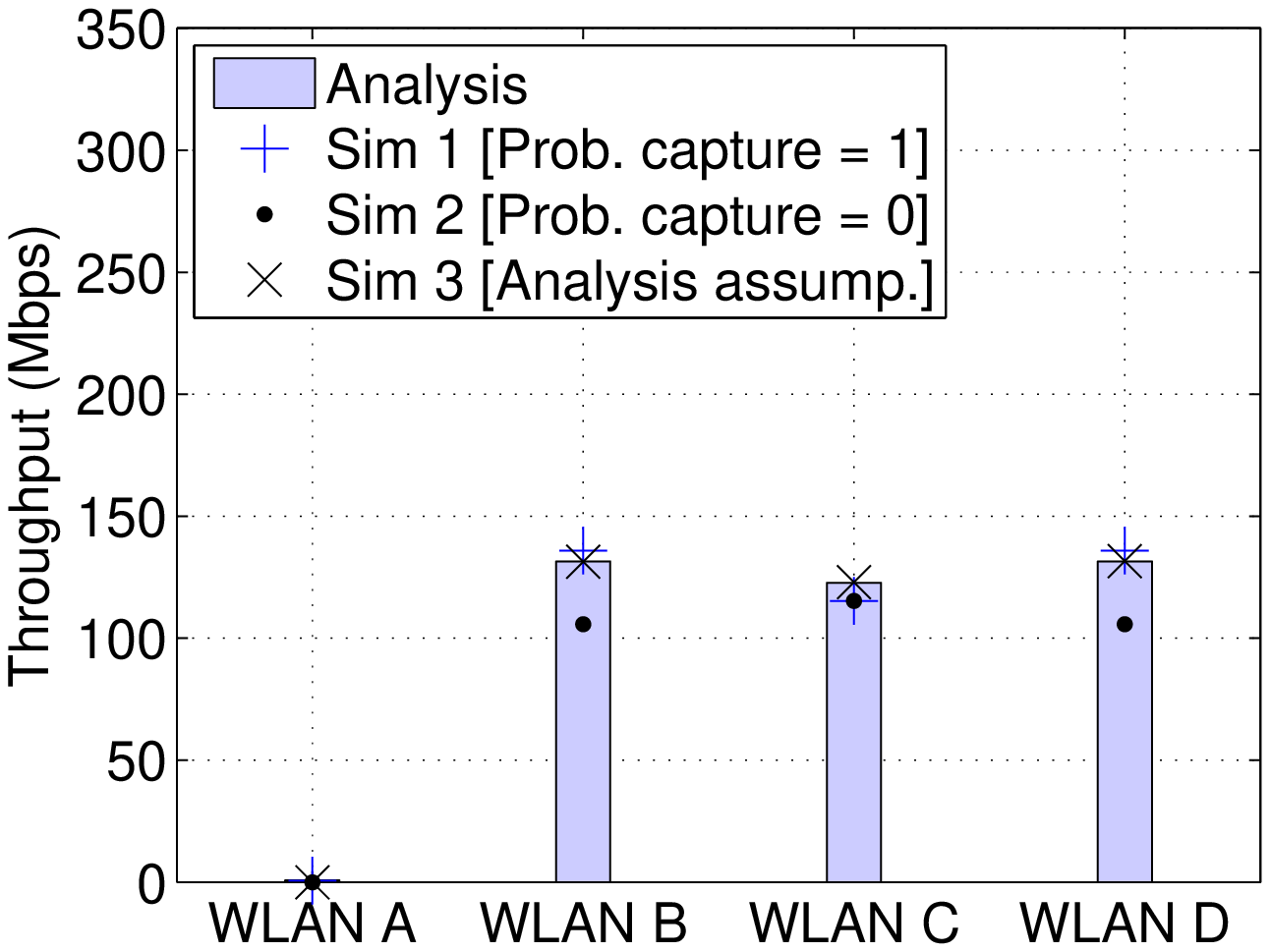,scale=0.32,angle=0}}
\caption{Throughput achieved by each WLAN when all of them have $2$ active nodes (i.e., the AP and one STA) in the four channel allocations considered. Each simulation result comes from a single simulation run of duration 10000 seconds.}\label{Fig:TN_ValSameN}
\end{figure}

\begin{figure}[ttth!]
\centering
\subfigure[Scenario 1]{\epsfig{file=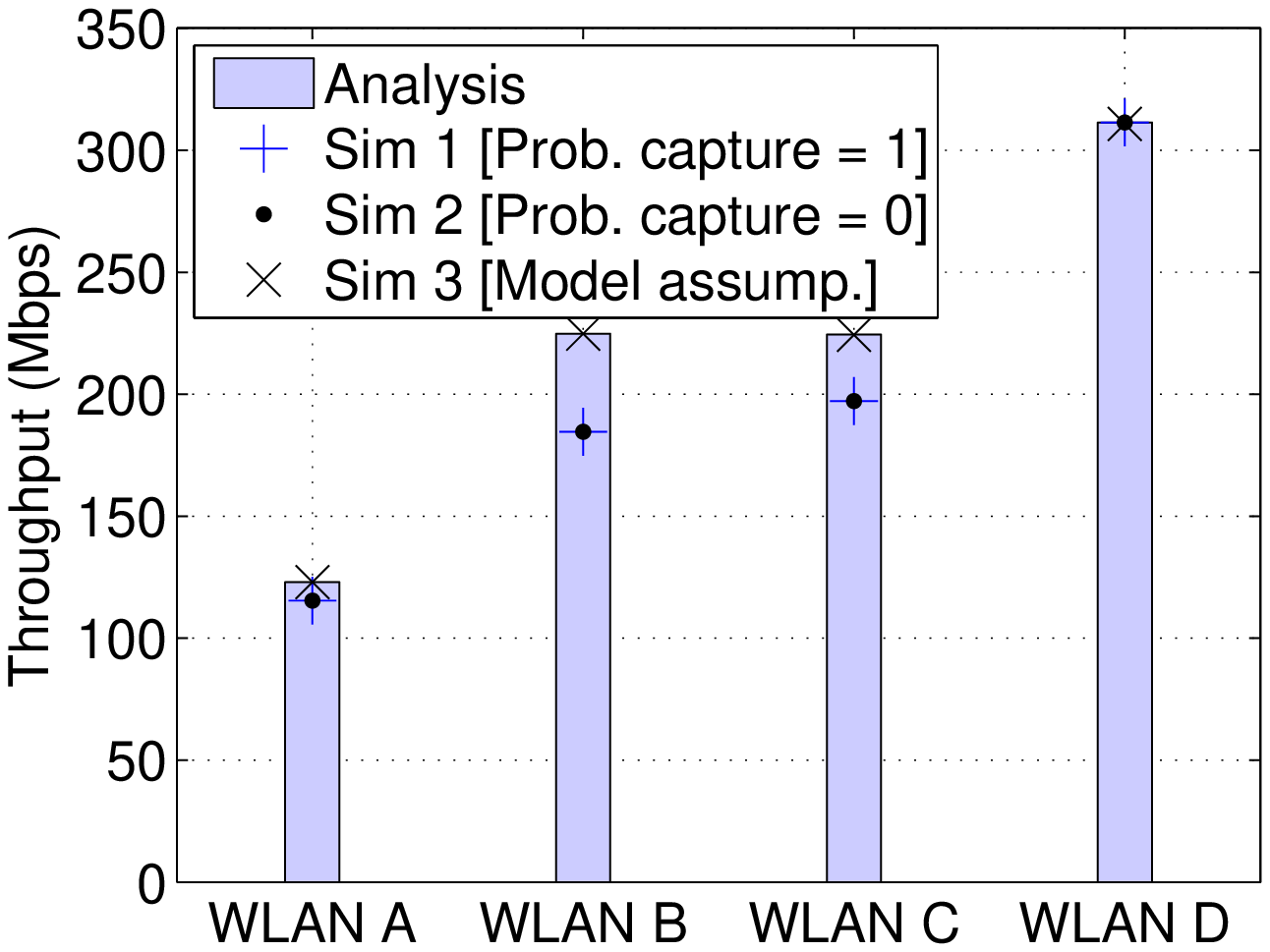,scale=0.32,angle=0}}\hspace{-0.5cm} 
\subfigure[Scenario 2]{\epsfig{file=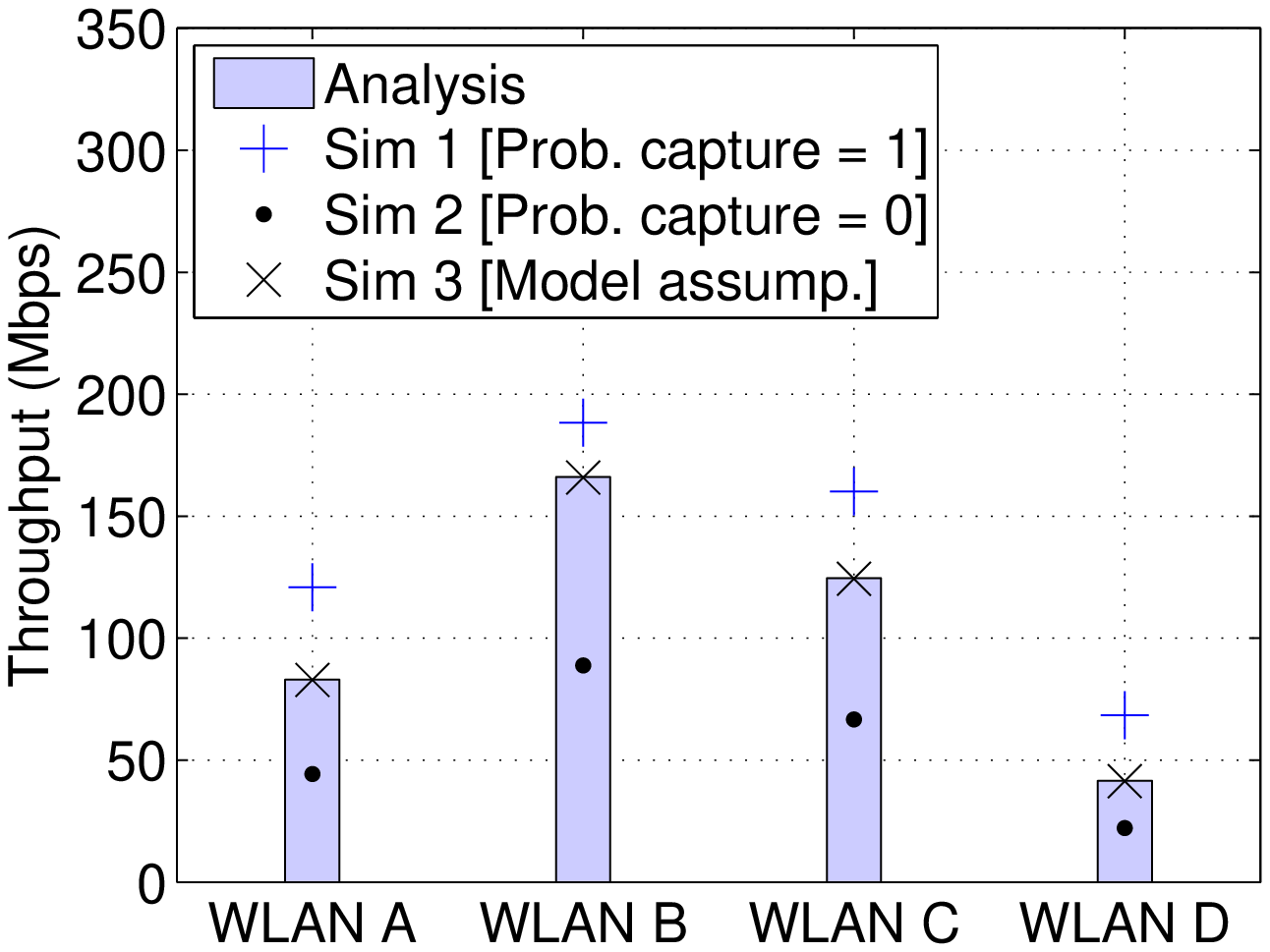,scale=0.32,angle=0}}\\ \vspace{-0.25cm}
\subfigure[Scenario 3]{\epsfig{file=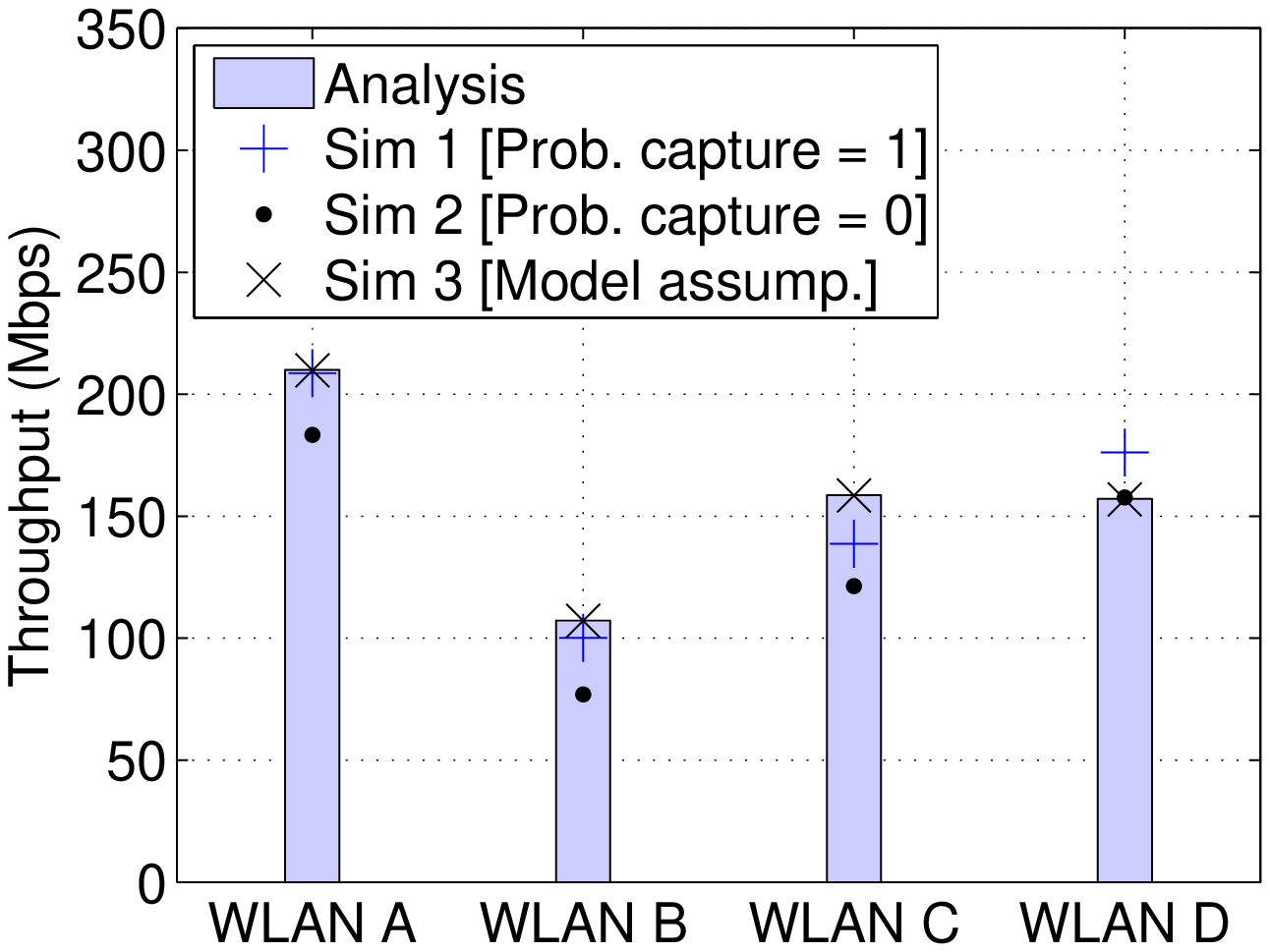,scale=0.32,angle=0}}\hspace{-0.5cm}
\subfigure[Scenario 4]{\epsfig{file=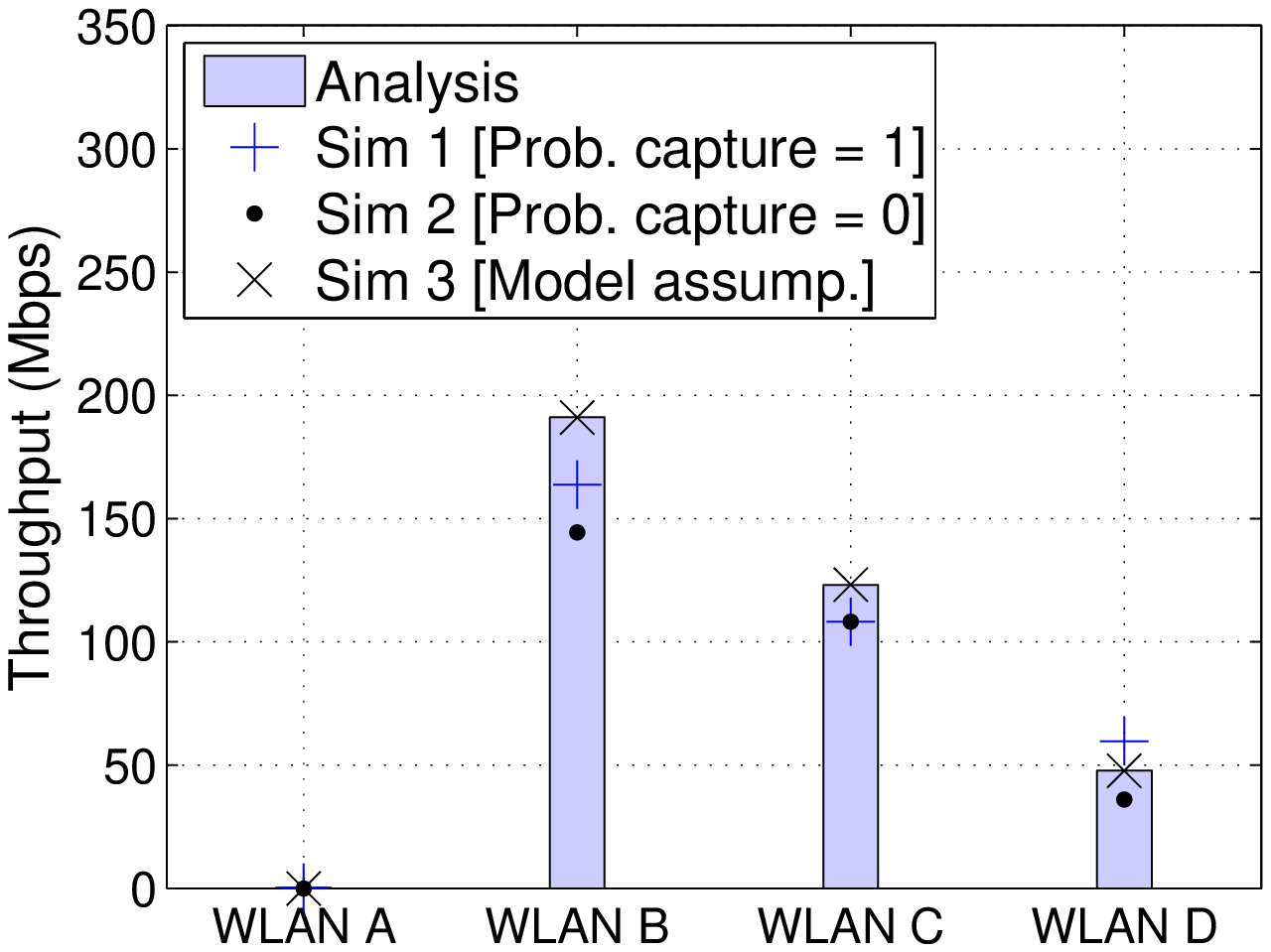,scale=0.32,angle=0}}
\caption{Throughput achieved by each WLAN in the four channel allocations considered when WLAN A has 1 active STA, WLAN B has $3$ active STAs, WLAN C has two active STAs and in WLAN D only the AP is transmitting packets. Each simulation result comes from a single simulation run of duration 10000 seconds.}\label{Fig:TN_ValDifN}
\end{figure}

The throughput achieved by each WLAN is plotted in Figure~\ref{Fig:TN_ValSameN} (all WLANs have two active nodes: the AP and one STA) and Figure~\ref{Fig:TN_ValDifN} (each WLAN has a different number of active STAs, exactly as shown in Figure~\ref{Fig:ToyNetworkBasic}). Comparing these two cases allows us to visualise the effect of a different number of active STAs in each WLAN on the system performance and to determine if modelling the aggregated operation of a WLAN instead of the operation of every node is a valid approach.

The throughput for the scenario in which all WLANs have the same number of nodes, and the scenario in which they do not are shown in Figures~\ref{Fig:TN_ValSameN} and~\ref{Fig:TN_ValDifN}, respectively. Four curves are plotted for each WLAN: the throughput computed using the WLAN-centric analytical model (bars), the throughput obtained from the simulator when the capture effect is considered (Sim 1), the throughput from the simulator when capture effect is not considered (Sim 2), and the throughput from the simulator when the same assumptions used for the analysis are considered (Sim 3).  

The results of the throughput model and Sim 3 match perfectly, which validates again the correctness of the results and shows that the insensitivity property indeed holds. Since the throughput model does not allow two or more nodes to transmit simultaneously, it does not benefit from concurrent packet receptions when the capture effect is enabled. Therefore, in some cases when the number of overlapping WLANs is high, the capture effect causes a higher throughput than the model (Sim 1). Otherwise, if packet capture is not considered, the achieved throughput is lower than the predicted by the analytical model due to the negative effect of collisions (Sim 2). The impact of each of the four channel allocations is discussed next.

Figure~\ref{Fig:TN_ValSameN} shows the throughput achieved by each WLAN in the four scenarios. In Scenario 1, the WLANs do not overlap because they use different groups of basic channels. Therefore, the throughput achieved by each WLAN only depends on the number of basic channels it uses. In Scenario 2, all WLANs overlap because they all use $8$ basic channels. In this case, all WLANs compete with all of the others for the channel, which results in the same throughput for all of them. A comparison of the results of {Scenarios 1 and 2} indicates that unless the packet capture effect is enabled, using {a single} basic channel is better than using $8$ basic channels if there is overlap with the other three neighboring WLANs. In Scenario 3, the channels of WLANs B and C are located between WLANs A and D, and they all use 4 basic channels. This situation benefits WLAN A and D because they only overlap with WLANs B and C, respectively, which are also competing for the channel resources. Lastly, Scenario 4 represents a random channel allocation. It is remarkable that WLAN A, which uses more basic channels, achieves nearly zero throughput. This occurs because it has to compete with the other three WLANs, which are in two independent groups that do not compete. WLANs B and D have the same throughput despite using different channel widths due to the performance anomaly.

Figure~\ref{Fig:TN_ValDifN} shows the throughput achieved by each WLAN in the same four scenarios as in Figure~\ref{Fig:TN_ValSameN} but with different numbers of active STAs in each WLAN. WLAN A has a single STA, WLAN B has three STAs, WLAN C has two STAs, and only the AP transmits in WLAN D. Increasing the number of STAs in a WLAN is equivalent to increasing its activity factor {$\theta$}, which also affects its throughput and how it interacts with the other networks. It is worth mentioning that a similar effect would be achieved by keeping the number of nodes per WLAN constant, but reducing the backoff duration.

In terms of fairness, we compute the JFI with respect to the throughput achieved by each WLAN. The results are shown in Table~\ref{Tbl:ValJFI}, where a low JFI value indicates that the four WLANs achieve very different throughputs.

\begin{table}
    \centering
    \begin{tabular}{ccc}
    \toprule
    \textbf{\bf Scenario} & \textbf{\bf Same Number of Nodes} & \textbf{\bf Different Number of Nodes} \\
    \midrule
    1 & 0.9135 & 0.91643 \\
    2 & 1 & 0.83333 \\
    3 & 0.90167 & 0.94987 \\
    4 & 0.75199 & 0.60826 \\     
    \bottomrule      
    \end{tabular}
    \caption{Jain's Fairness Index}\label{Tbl:ValJFI}
\end{table}

\begin{figure}[t!]
\centering
\epsfig{file=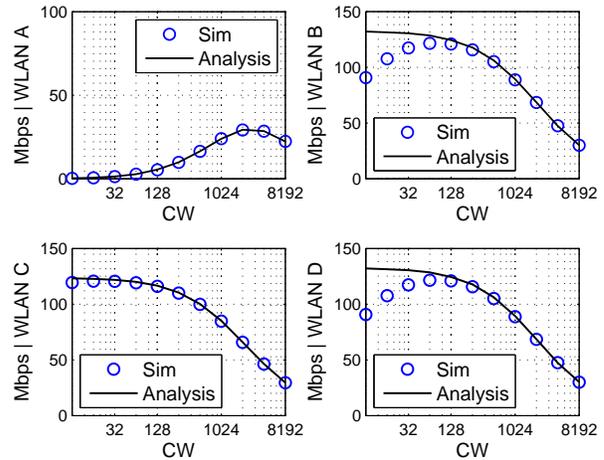,width=\columnwidth,angle=0}\\
\caption{Throughput achieved by each WLAN in Scenario 4 when each one has two active nodes. The probability of capturing a packet in case of collisions is set to $0$, and therefore we are considering the worst case in terms of the negative effect of collisions.}\label{Fig:CW4WLANs}
\end{figure}

Figure \ref{Fig:CW4WLANs} shows the throughput achieved by each one of the four WLANs in Scenario 4 when  the CW increases from $8$ to $8192$. We consider the case in which WLANs have two nodes active. The continuous time backoff mechanism is able to capture the same dynamics as when a discrete backoff mechanism (i.e., as in IEEE 802.11 WLANs) is considered. Only when the effect of collisions is significant, the continuous time backoff mechanism offers optimistic results. Moreover, it can be observed that almost exact values are achieved in both cases when the CW value is optimal for the discrete backoff scheme (i.e., the CW value that maximizes the throughput) since it is the value at which the negative effect of collisions becomes marginal. Besides that, Figure \ref{Fig:CW4WLANs} also shows that increasing the CW  value we can reduce the starvation suffered by WLAN A. The downside is that we reduce severely the throughput achieved by the other three WLANs.

\section{Channel Allocation Schemes} \label{Sec:ChannelAllocation}

Neighboring WLANs operating in the Industrial, Scientific and Medical (ISM) band may belong to different administrative domains, and therefore they may select the channel to use autonomously and, in most of the cases, without any information about the current spectrum occupancy. This situation is equivalent to select the channel to use uniformly at random by each WLAN.

In this section, we describe such random channel selection approach, considering two channelisation cases: $i$) any group of basic channels can be selected, and $ii$) only the channels specified by the IEEE 802.11ac amendment can be selected. In order to determine the network capacity that is lost because of the absence of a controlled channel allocation, we also introduce an optimal centralised proportional fair channel allocation strategy.

\subsection{Decentralised approaches}

Channel allocation in autonomous WLANs is done in a decentralised way. That is, each WLAN chooses the group of basic channels to use independently. In this category we consider two cases:

\subsubsection{Random Channel Selection}

In this scheme, WLAN $i$ uniformly selects $c_i$ consecutive basic channels at random from the $N$ available basic channels. 

\subsubsection{IEEE 802.11ac channelisation}

IEEE 802.11ac channelisation tries to prevent that WLANs using the same number of basic channels partially overlap. This is achieved by explicitly defining the groups of basic channels that can be selected when $c_i$ channels are going to be used. Namely, given that a WLAN is going to use $c_i$ basic channels, it can only select $\lfloor \frac{N}{c_i} \rfloor$ different channels. Once one of these channels is selected, the first basic channel in it is $c_i (Z_1-1)+1$, and the last one is $c_i Z_1$, where $Z_1=\mathcal{U}([1,\ldots,\lfloor\frac{N}{c_i} \rfloor )$ is an uniformly distributed random value between $1$ and $\lfloor\frac{N}{c_i} \rfloor$. Note that the available basic channels are numbered from $1$ to $N$. 

\subsection{Centralised approach}

When all WLANs are mutually within carrier sense range, we characterise the optimal \emph{proportional fair channel allocation} as a linear combination of \emph{waterfilling solutions}, assuming that WLANs can alternate periodically between different allocations. This optimal allocation is the best trade-off between maximising throughput and fairness, in the sense that, starting from the optimal allocation, a proportional increase of the throughput for any set of WLANs would result in a bigger proportional decrease of throughput for the remaining WLANs.

Then, we will show how to relax the two assumptions we made in the following way:
\begin{itemize}
\item when not all WLANs are mutually within carrier sense range, we present a technique to devise a sub-optimal solution;
\item if the WLANs cannot alternate periodically between different allocations, we show that a single waterfilling solution is a reasonable sub-optimal choice.
\end{itemize}

This is an idealised approach, where a central server with knowledge of the WLANs topology is needed. However, the computation required only depends on the number of WLANs that mutually interfere and on the number of basic channels available. Moreover it is easy to compute in an efficient way, and such computation can be done preemptively.

\subsubsection{Proportional Fair Channel Allocation}  \label{sec:prop-fairn}

Let $\field{K}$ be the collection of all possible sets of channels, \ie $\field{C} \in \field{K}$. We call  $x(\field{C})$ the corresponding aggregate throughput of a set. For example, each of the four channel allocations represented in Figure~\ref{Fig:ChannelAssignment} has a different $\field{C} \in \field{K}$ and a corresponding aggregate throughput $x(\field{C})$. 

We want to characterise the optimal $\field{C} \in \field{K}$. However, this problem is in general hard to solve because of the combinatorial structure of the discrete collection of sets $\field{K}$. To simplify the analysis, we need to allow WLANs to switch their channel configuration $\field{C}$ at any time and look for an optimal time schedule for the network along a time period. In other words, we allow WLANs to switch to a different $\field{C}$ and keep that configuration for a certain period of time. 

We define a \emph{(global) schedule} $p(\field{C}) \colon \field{K}\mapsto \left[0,1\right]$ as the portion of time the network spends on each channel configuration $\field{C}$. Since we are including all the possible channel configurations in $\field{K}$, including those in which some WLANs are not transmitting at all (\ie $C_i = \emptyset$), the schedule vector must sum to one, i.e.,  $\sum_{\field{C} \in \field{K} } p(\field{C}) = 1$.

For example, considering Figure~\ref{Fig:ChannelAssignment}, a possible (although clearly not optimal) schedule would be the one in which the system uses Scenario 1 for half of a time period, and no WLAN is transmitting ($C_i = \emptyset$ for all $i$) for the rest of the time.

To determine the proportional fair global scheduling, we need to solve the following utility optimisation problem:

\begin{problem}[Proportional Fairness]\label{prob:1}
\begin{align*}
&\max_{p} \sum_{i=1}^M \log \sum_{ \field{C} \in \field{K}} p(\field{C}) x_i(\field{C}) \\
\text{s.t.} &\sum_{\field{C} \in \field{K} } p(\field{C}) = 1,\\
& p(\field{C}) \in \left[0,1\right], \quad \text{for all $\field{C} \in \field{K}$}.
\end{align*}
\end{problem}

The quantity $\sum_{ \field{C} \in \field{K}} p(\field{C}) x_i(\field{C})$ is the throughput achieved by WLAN $i$ using the schedule $p(\field{C})$, and it is computed as the weighted average of the different throughputs $x_i(\mathcal C)$ for the various  $\mathcal C \in \mathcal K$.

\paragraph{Properties of Problem~\ref{prob:1}} 

This problem requires the maximisation of a concave function in a convex set; thus, it is easy to solve in principle. The objective function is concave because when $p(\field{C})$ is a vector with $|\mathcal K|$ entries, $f_i(p(\field{C})) = \sum_{\field{C} \in \field{K}} p(\field{C}) x_i(\field{C})$ is affine, so $\log{f_i(p(\field{C}))}$ is concave because is composed of an affine function, and the sum of concave functions is concave. Moreover, the constraints are clearly convex. This formulation is broad enough to include the case when not all WLANs are mutually within carrier sense range.

Unfortunately, the size of $\field{K}$ grows exponentially with $M$, which makes the computation of the throughput function $x(\cdot)$ challenging. To overcome this issue we will now characterise more in detail the optimal and sub-optimal solutions, to be able to derive them without explicitly solving the convex problem.

\paragraph{Waterfilling Solution} 

We can define the waterfilling solution when all WLANs are mutually within carrier sense range. We will relax this assumption later. Given the number of basic channels $N$, we can easily build a mapping from the number of WLANs $M$ to the allocation that minimises the number of overlaps between WLANs.

Algorithm~\ref{algo:waterfilling} shows the pseudo-code to build such a mapping, $f(M,N)$ (similarly as~\cite{radunovic2002unified} for the case of free-disposal property). The number of basic channels used is doubled once for each WLAN until the number of available basic channels allows it. The first channel positions are then chosen such that the spectrum is evenly used. 

\begin{algorithm}[t]
\begin{algorithmic}[1]
	\State{assign to each WLAN a single basic channel, \ie $c_i = 1$ for all $i=1,\dots,M$.}
        \Loop{}
          \For{$i=1,\dots,M$}
          \If{$2c_i + \sum_{j\neq i} c_j \le N$}
            \State{$c_i \leftarrow 2c_i$}
          \Else
            \State{\textbf{goto} 11}
          \EndIf
          \EndFor
        \EndLoop
        \State{For each WLAN $i$, select the basic channels as the contiguous set $[1+\sum_{j<i}{\min(c_j,W_{\max})},\sum_{j\le i}{\min(c_j,W_{\max})}]$  modulo $N$.}
\end{algorithmic}
\caption{Waterfilling Algorithm}
\label{algo:waterfilling}
\end{algorithm}

This procedure always produces an allocation that minimises the number of overlaps per channel. Moreover, the obtained allocation is such that either all WLANs have the same width, or there are only two sets of widths. In the latter case we can split the WLANs into two sets $G_1$ and $G_2$, such that $c_i = 2\cdot c_j$ for each $i\in G_1, j\in G_2$.

Such waterfilling allocation plays a key role in the proportional fair allocation, even in the relaxed cases of when not all WLANs are mutually within carrier sense range and when WLANs cannot alternate periodically between different allocations, we show in the following that a single waterfilling solution is a good sub-optimal choice.

\paragraph{Proportional Fairness and Waterfilling} 

We now present a conjecture regarding the relationship between the waterfilling configuration and the proportional fair configuration.

\begin{conjecture} \label{conj:prop}
 The proportional fair solution to Problem~\ref{prob:1} with the throughput function that was defined in Section~\ref{Sec:ThroughputModel} is a linear combination on waterfilling configurations only. 
\end{conjecture}

This means that to have a proportional fair configuration, the WLANs should change roles in turn between the (non-unique) waterfilling solutions.

If such a time slicing function is not available, any solution from the waterfilling configuration is an acceptable sub-optimal solution. We corroborate this claim by means of simulation in Section~\ref{Sec:NumResults}, see  Figure~\ref{Fig:PropFair} in particular.

We simulated different scenarios, and solved Problem~\ref{prob:1} using the Matlab CVX framework. Conjecture~\ref{conj:prop} was never confuted in our simulations.

\paragraph{Interactions between multiple groups of neighboring WLANs}

We present a technique to devise a sub-optimal solution when not all WLANs are mutually within carrier sense range.  We need such a technique because, although Problem~\ref{prob:1} would still represent such scenarios and would still be convex, Conjecture~\ref{conj:prop} is not valid anymore. Consequently, characterising the optimal solution becomes very hard in general.

First we need to consider the interference graph of the network $G=(V,E)$, where  $V$ is the set of WLANs, and the edges are defined as $e=(i,j)\in E$ if WLAN $i$ can interfere with WLAN $j$. Interference is assumed to be symmetric and thus $G$ is an undirected graph.

We can  compute the chromatic number $\chi$ of this graph, i.e., the minimal number of colors necessary to have the property that no neighbours share the same color. A coloring with $\chi$ colors represents an equivalence relation of minimum cardinality such that all WLANs that share the same color do not interfere, and thus can choose the same set of channels.

Therefore, we can consider the collection of $\chi$ groups of WLANs with same color as a collection of virtual WLANs, and use the mapping $f(\chi,N)$ obtained using Algorithm~\ref{algo:waterfilling} over these virtual WLANs, i.e., we use a waterfilling allocation where all WLANs that share same color will have the same allocation.

\begin{figure}[t!]
\centering
\psfrag{WLAN A}[][][0.7]{WLAN A}
\psfrag{WLAN B}[][][0.7]{WLAN B}
\psfrag{WLAN C}[][][0.7]{WLAN C}
\psfrag{WLAN D}[][][0.7]{WLAN D}
\psfrag{WLAN E}[][][0.7]{WLAN E}
\psfrag{WLAN F}[][][0.7]{WLAN F}
\psfrag{WLAN G}[][][0.7]{WLAN G}
\psfrag{WLAN H}[][][0.7]{WLAN H}
\epsfig{file=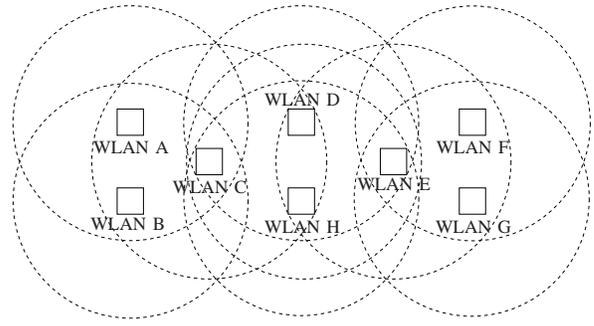,scale=0.35,angle=0}\\
\caption{Eight WLANs distributed in four groups of neighboring WLANs: WLANs A, B and C in group 1; WLANs C, D and H in group 2; WLANs D, E and H in group 3; and WLANs E, F and G in group 4.}\label{Fig:SpatDisScenario}
\end{figure}

As an example, let us consider the scenario depicted in Figure~\ref{Fig:SpatDisScenario}. The number of available basic channels is set to $N=19$. WLANs select both the width and the position of the selected channel uniformly at random. The set of available channel widths is $\{20,~40,~80,~160\}$ MHz. 

\begin{figure}[t!]
\centering
\epsfig{file=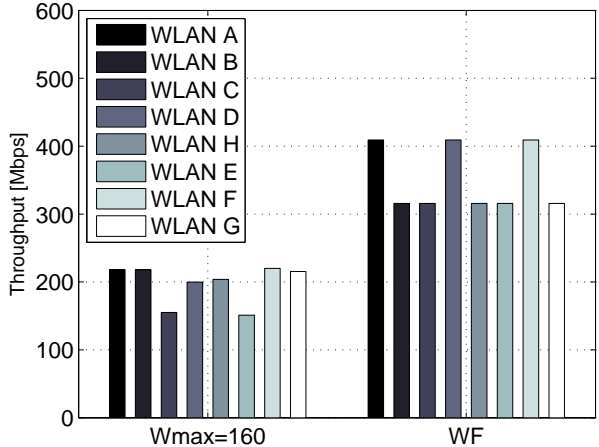,width=\columnwidth,angle=0}  
\caption{Expected throughput achieved by each WLAN in the scenario shown in Figure \ref{Fig:SpatDisScenario}. Random channel allocation versus waterfilling algorithm.}\label{Fig:MCWN19}
\end{figure}

In this case we have $\chi=3$ and the groups that share the same color are $\{I_1,I_2,I_3\}=\{\{A,D,F\},\{C,E\},\{B,H,G\}\}$. If we run the waterfilling algorithm on $\{I_1,I_2,I_3\}$ we get the same solution of a complete graph with three WLANs, so one WLAN will have width equal to 8 and two WLANs will have width equal to 4. A possible solution obtained with this technique is the following:
$C_A=\{1-8\}$, $C_B=\{13-16\}$, $C_C=\{9-12\}$, $C_D=\{1-8\}$, $C_H=\{13-16\}$, $C_E=\{9-12\}$, $C_F=\{1-8\}$, and $C_G=\{13-16\}$, which as shown in Figure \ref{Fig:MCWN19} results in a higher throughput than using the random channel allocation scheme. In case of all WLANs are interfering with each other, then the set $I$ corresponds simply with the set of WLANs.

If WLANs can alternate between channel allocations, then all WLANs will have on average the same throughput, alternating the roles amongst the color groups. But even in this case the solution is in general sub-optimal, because a very unbalanced interference graph could allow more aggressive solutions, i.e., it could happen that some nodes belonging to a certain color group would be allowed to select a wider width than the other nodes in the same group.


\section{Performance Evaluation} \label{Sec:NumResults}
 
In this section, we evaluate the impact to the system of a different number of neighboring WLANs, a different number of available basic channels and the set of channel widths that are available for each WLAN. 

\subsection{Increasing the number of channels}

\begin{figure*}[t!]
\centering
\subfigure[Spectrum Utilization]{\epsfig{file=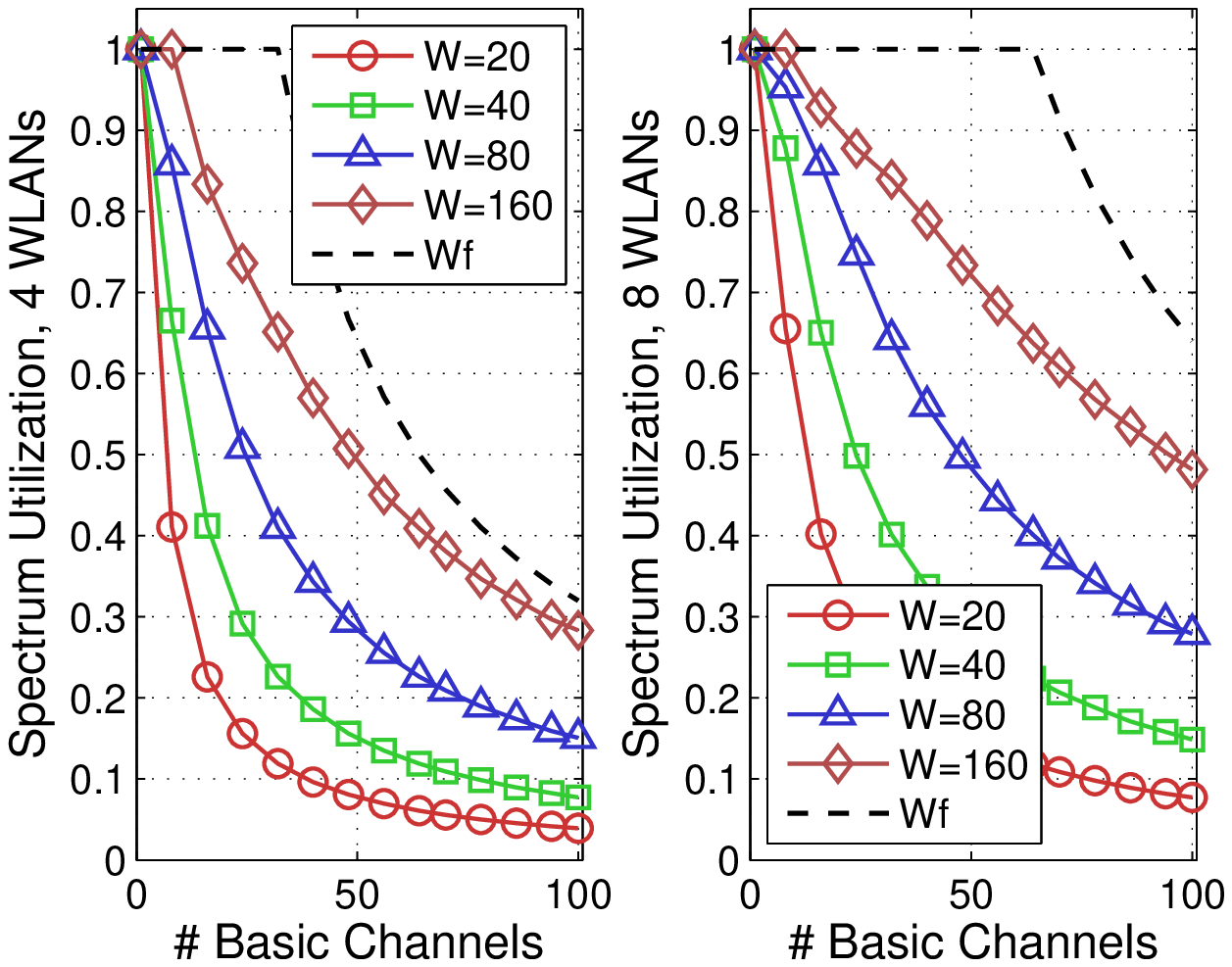,scale=0.44,angle=0}\label{Fig:C_VC_fix}}\hspace{-0.75cm} 
\subfigure[Expected Throughput per WLAN]{\epsfig{file=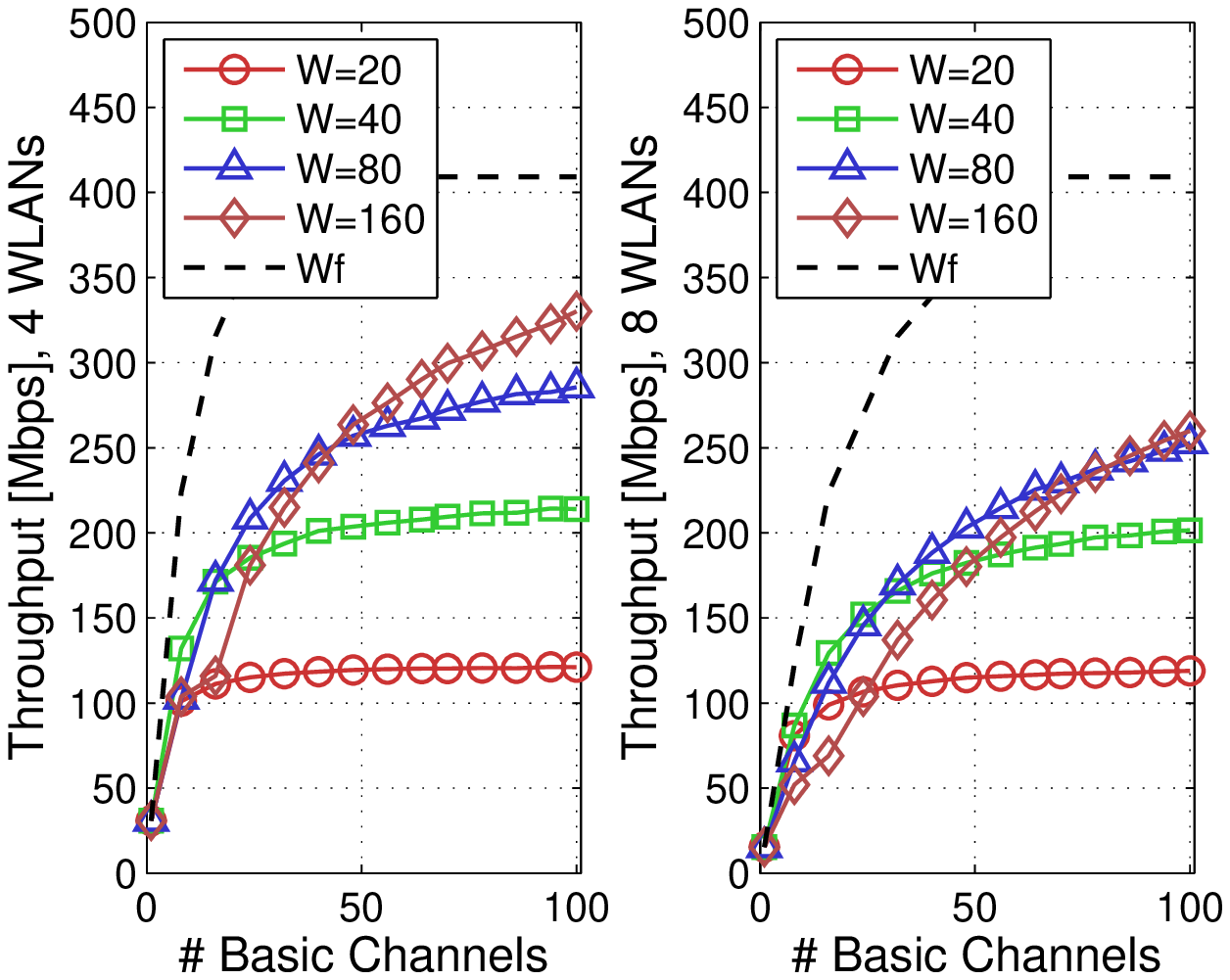,scale=0.44,angle=0}\label{Fig:S_VC_fix}}  \hspace{-0.75cm} 
\subfigure[Jain's Fairness Index]{\epsfig{file=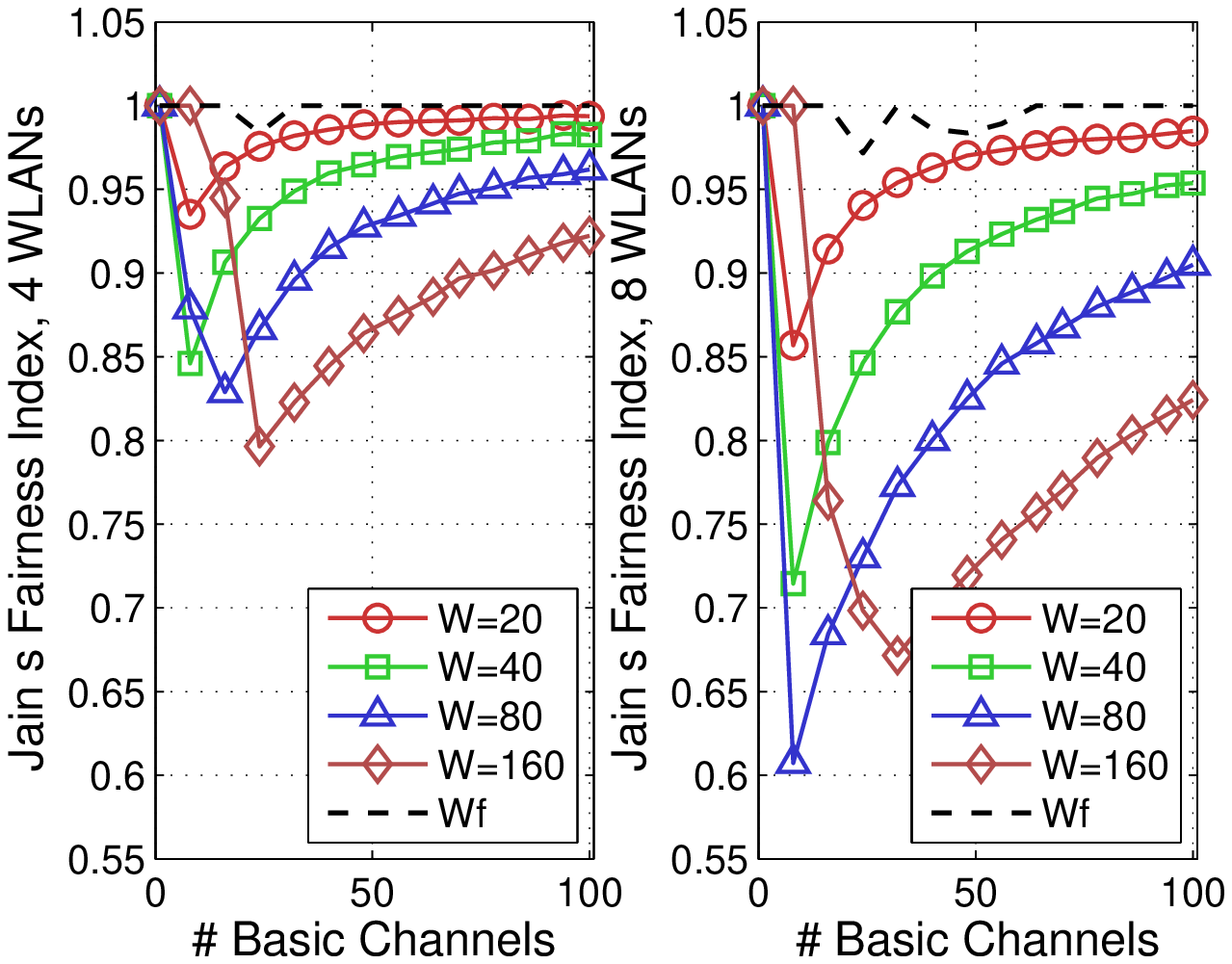,scale=0.44,angle=0}\label{Fig:F_VC_fix}} 
\caption{Throughput, Spectrum Utilization and Fairness when the number of basic channels increases.}\label{Fig:ResultsIncChannels}
\end{figure*}

Figure~\ref{Fig:ResultsIncChannels} shows the expected spectrum utilisation (Figure~\ref{Fig:C_VC_fix}), the expected throughput of a single WLAN (Figure~\ref{Fig:S_VC_fix}) and the expected throughput fairness (Figure~\ref{Fig:F_VC_fix}) for $4$ and $8$ neighboring WLANs when the number of basic channels increases from $1$ to $100$ and all WLANs use the same channel width $W$. The spectrum utilisation is computed as the fraction of basic channels that are occupied by one or more WLANs versus the total number of basic channels, i.e.,
    	\begin{align}\label{Eq:ChannelUtilization}
            v(\mathcal{C}) :=\frac{1}{N}\sum_{k=1}^{N}{I(k)}. 
     \end{align}
     where the function $I(k)$ returns $1$ if the basic channel $k$ is found occupied by one or more WLANs.

The results from Figure~\ref{Fig:ResultsIncChannels} show that the waterfilling algorithm is able to maximise the spectrum utilisation while distributing the available basic channels evenly among the neighboring WLANs. As a consequence, it provides the highest WLAN network throughput and fairness. The JFI values that are less than 1 are obtained when not all WLANs in the resulting channel allocation have allocated the same channel width.

When each WLAN randomizes the group of basic channels to be used without any information about the spectrum occupancy or the number of neighbours, selecting a large group of basic channels only guarantees a higher throughput when the number of neighboring WLANs is small or when the number of available basic channels is very large. For example, when there are only four neighboring WLANs, selecting $W=160$ MHz only gives a higher throughput than $W=80$ MHz if more than $50$ basic channels are available. Finally, in terms of fairness, the use of a large $W$ also accentuates the differences in the throughput achieved by each WLAN. The fairness is therefore low because most of the neighboring WLANs overlap, which results in a significantly lower throughput than the few that do not.

\begin{figure}[h!]
\centering
\subfigure[$N=8$, $M=6$. The expected throughput values are: $89.847$, $103.21$, $78.965$, $69.031$ Mbps for $W=1$, $2$, $4$, and $8$ respectively.]{\epsfig{file=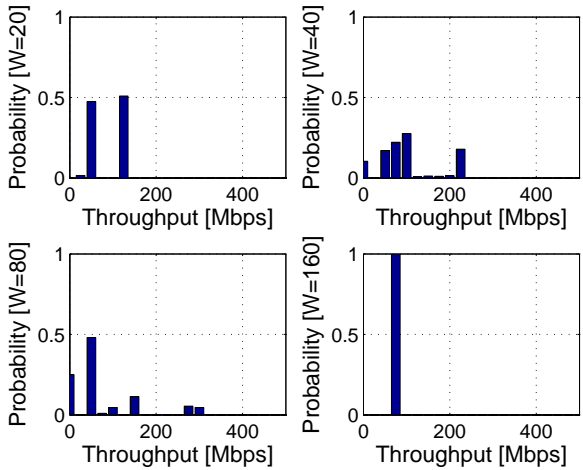,width=\columnwidth,angle=0}\label{Fig:HistW6N8}} 
\subfigure[$N=24$, $M=6$. The expected throughput values are: $110.23 $, $166.28$, $170.33$, $130.7$ Mbps for $W=1$, $2$, $4$, and $8$ respectively.]{\epsfig{file=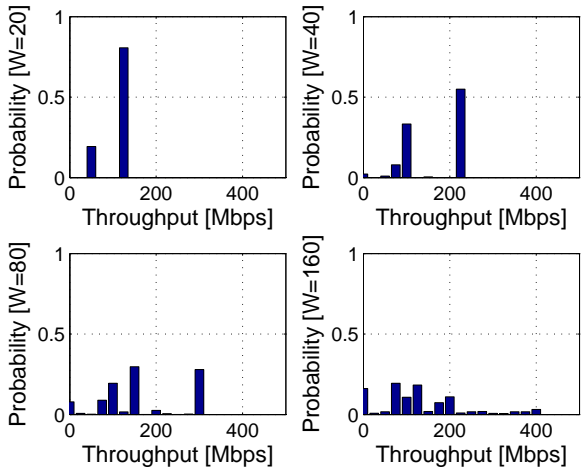,width=\columnwidth,angle=0}\label{Fig:HistW6N24}} 
\caption{Histogram of the aggregate throughput for different $N$ and $M$ values}\label{Fig:ResultsHist}
\end{figure}

To obtain more insight into the system dynamics, Figure~\ref{Fig:ResultsHist} shows the histogram of the achieved throughput by a single WLAN when there are $6$ neighboring WLANs and two numbers of basic channels: $N = 8$ and $N = 24$. We used the throughput of $10000$ randomly generated scenarios to obtain these histograms. The histogram shows all of the possible throughput values and the probability of achieving each one. For $N = 8$ (Figure~\ref{Fig:HistW6N8}) and $W = 20$ MHz, the throughput achieved by a single WLAN is higher than $100$ Mbps in approximately 50\% of the cases, which corresponds to the case in which none of the WLANs overlap. Increasing the channel width to $W = 40$ MHz increases the chances that the WLANs overlap, which reduces the expected WLAN throughput. However, in approximately 20\% of the cases, the WLANs randomly select $40$ MHz non-overlapping channels, which results in a higher throughput. Similar observations can be made for $W = 80$ MHz, where a maximum throughput of approximately $300$ Mbps can be achieved in only a few cases, as it is more likely to obtain a lower throughput than when using $W = 40$ MHz due to the higher overlapping probability. Finally, there is a single throughput value for $W = 160$ MHz because all WLANs overlap. Similar observations can be made for $N = 24$ basic channels (Figure~\ref{Fig:HistW6N24}). In this case, it is clear that the presence of more basic channels allows more combinations and improves the overall system performance. In this example, the optimal value of $W$, which is the $W$ value that results in the highest average throughput, is $40$ MHz ($103.21$ Mbps) and $80$ MHz ($170.33$ Mbps) for $N = 8$ and $N = 24$, respectively. Note that the expected WLAN throughput achieved in each case is shown in the caption of Figure~\ref{Fig:ResultsHist}.

\subsection{Increasing the number of WLANs}

\begin{figure*}[tH!]
\centering
\subfigure[Spectrum Utilisation]{\epsfig{file=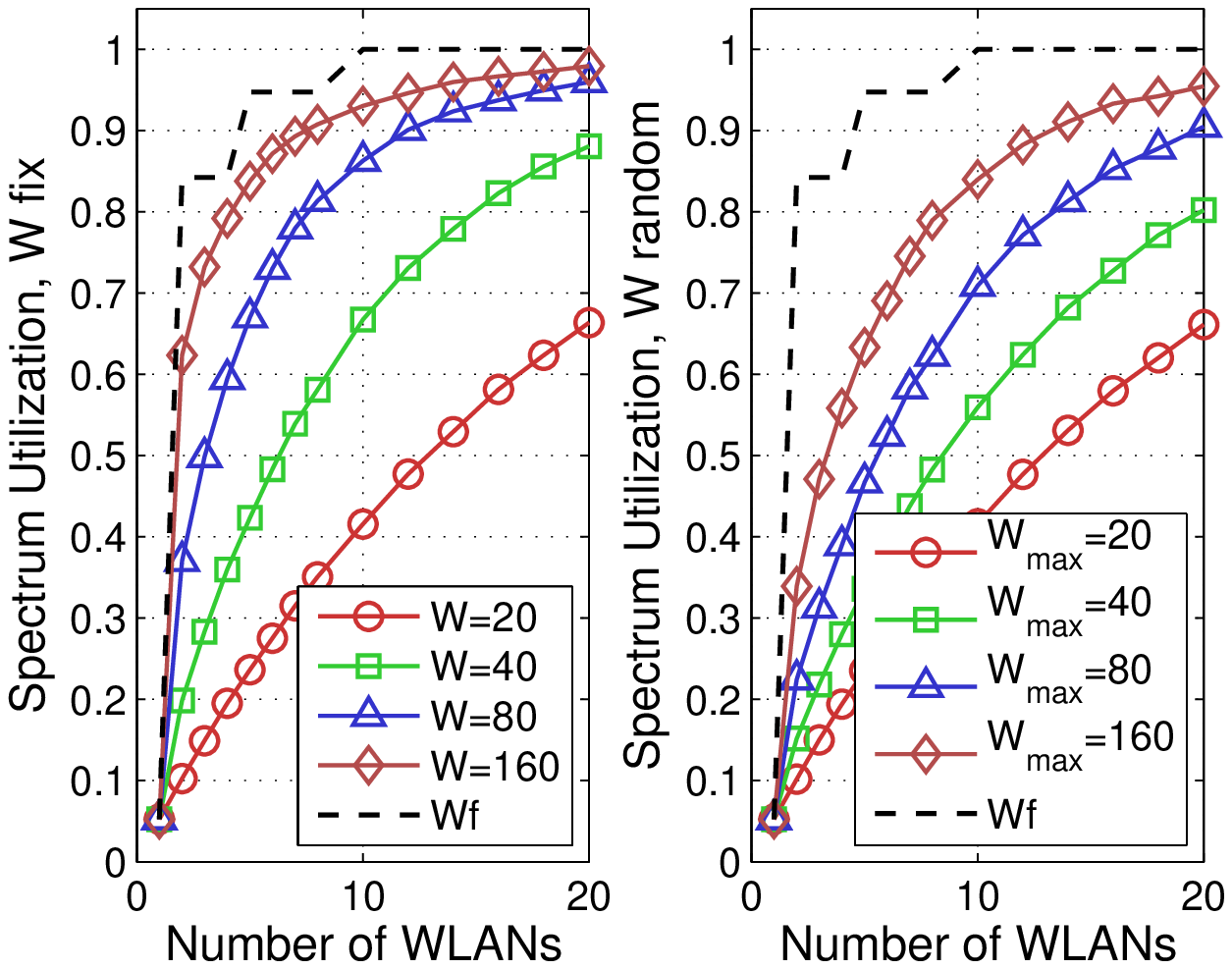,scale=0.44,angle=0}\label{Fig:C_VW_fix_rnd}} \hspace{-0.75cm}
\subfigure[Expected Throughput per WLAN]{\epsfig{file=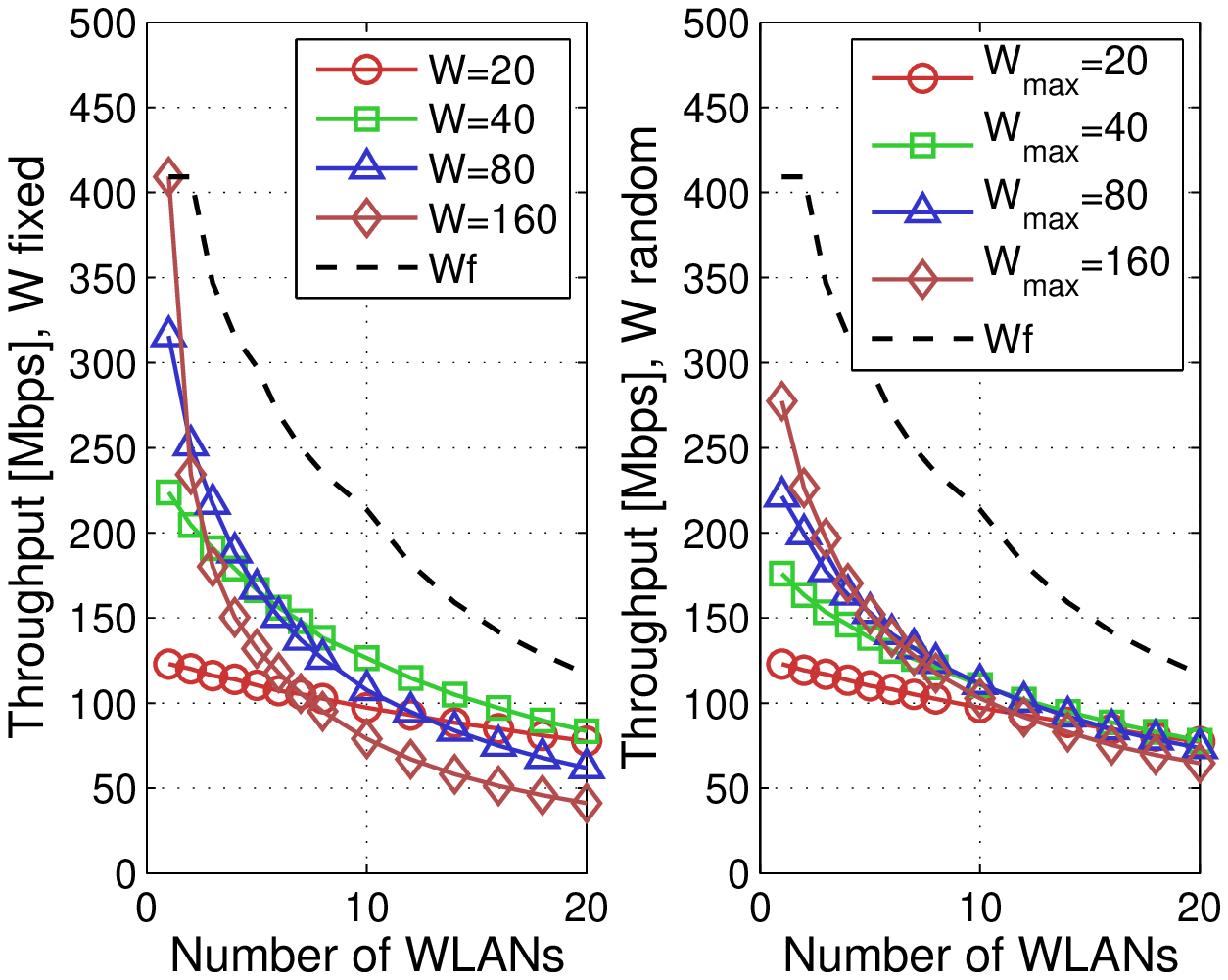,scale=0.44,angle=0}\label{Fig:S_VW_fix_rnd}} \hspace{-0.75cm} 
\subfigure[Jain's Fairness Index]{\epsfig{file=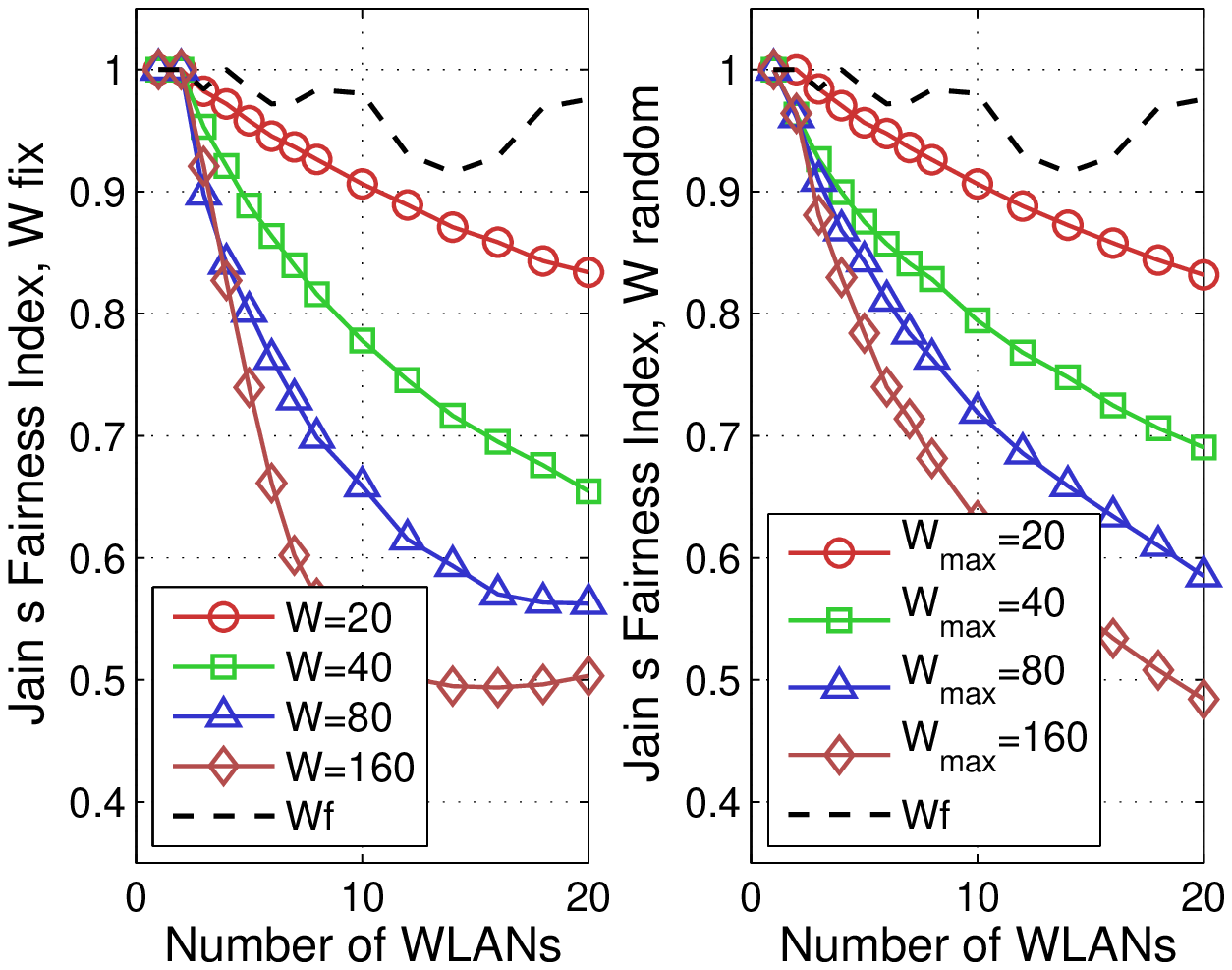,scale=0.44,angle=0}\label{Fig:F_VW_fix_rnd}} 
\caption{Throughput, Spectrum Utilisation and Fairness when the number of WLANs increases.}\label{Fig:ResultsInWLANs}
\end{figure*}

In Figure \ref{Fig:ResultsInWLANs} we show the system performance when the number of neighboring WLANs increases. The number of available basic channels is set to $N=19$. We also evaluate the case in which each WLAN randomly chooses the value of $W$ given a maximum value, $W_{\max}$ (i.e., $W$ is a random value that is uniformly distributed between the feasible values of $20$ Mhz and $W_{\max}$). 

The results show that increasing the number of WLANs results in a higher spectrum utilisation (Figure \ref{Fig:C_VW_fix_rnd}), lower throughput (Figure 12(b)) and generally lower fairness (Figure \ref{Fig:F_VW_fix_rnd}). Note that the effect of randomly selecting $W$ increases the ways that different WLANs interact because more combinations are feasible, which for any $W_{\max}$ value and a large number of WLANs results in a throughput similar to the achieved when $W_{\max}=20$ MHz. However, the fairness decreases with $W_{\max}$. 


\begin{figure}[h!!!!!]
\centering
\epsfig{file=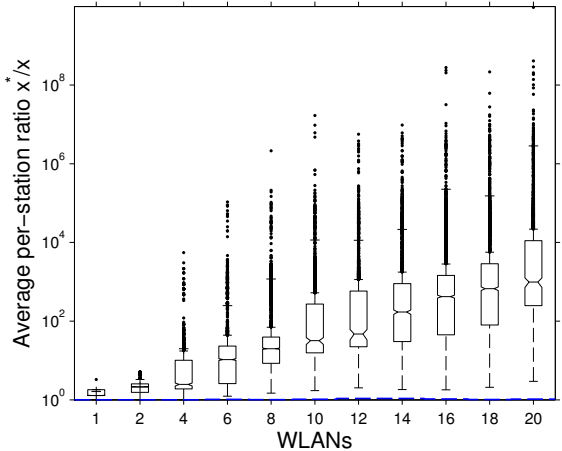,scale=0.6,angle=0}\\
\caption{Average per-WLAN ratio distribution $\frac{1}{M}\sum_{i=1}^M \frac{x_i^*}{x_i}$ between optimal throughput and sampled throughput of random channel allocations with $W_{\max}=160$\,MHz. Blue dotted line represents the waterfilling solution.  }\label{Fig:PropFair}
\end{figure}

Figure~\ref{Fig:PropFair} shows the average per-WLAN ratio distribution $\frac{1}{M}\sum_{i=1}^M \frac{x_i^*}{x_i}$ between the optimal throughput and the sampled throughput of random allocations with $W_{\max}=160$\,MHz. The same quantity for the waterfilling solution is shown. The box represents the samples inside the interquartile
range $Q3-Q1$, the crosses represent the average and the notches represent the medians. The black dots represents outliers (samples more than $1.5$ times the interquartile range). Any solution that is different than the proportional fair solution makes the sum of the proportional gain negative (and thus also the average). When $W_{\max}=160$\,MHz, rare events of starvation effect occur, especially when the number of WLANs increase.

\subsection{IEEE 802.11ac channelisation}

In Table \ref{Tbl:11acChannels} we show the expected aggregate throughput and throughput fairness for the two decentralised channel allocation schemes considered in this work. The considered number of available basic channels is set to $N=16$ for a fair comparison between both schemes. Similar results between both channel selection schemes are obtained. However, since the IEEE 802.11ac channelisation prevents the negative effects of partial overlaps between WLANs, and non-direct interactions as well, it results in a slightly better aggregate throughput and throughput fairness for large $W$.

\begin{table*}[t!!]
    \centering
    \begin{tabular}{ccccccccccc}
        \toprule
         \multicolumn{3}{c}{\bf Parameters}  & \multicolumn{4}{c}{\bf Aggregate Throughput [Mbps]} & \multicolumn{4}{c}{\bf JFI} \\  
        \midrule
         &  &  & \multicolumn{4}{c}{$W_{\max}$ [MHz]} & \multicolumn{4}{c}{$W_{\max}$ [MHz]} \\
        Channelisation & N & M & 20 & 40 & 80 & 160  & 20 & 40 & 80 & 160 \\
        \hline                 
        Random & 16 & 8 & 789.1 & 897.6 & 909.2 & 844.0 &  0.95 & 0.95 & 0.93 & 0.91 \\        
        802.11ac & 16 & 8 & 794.2 & 936.4 & 966.5 & 928.2 & 0.95 & 0.95 & 0.95 & 0.93 \\
        \hline 
        Random & 16 & 12 & 1058.4 & 1119.7 & 1092.9 &  974.0 &  0.96 & 0.95 & 0.92 & 0.89  \\        
        802.11ac & 16 & 12 & 1058.4 & 1156.6 & 1157.1 & 1081.2 & 0.96 & 0.96 & 0.94 & 0.92 \\
        \hline
        Random & 16 & 16 & 1264.7 & 1276.7 & 1212.2 & 1065.3 & 0.97 & 0.94 & 0.91 & 0.87 \\        
        802.11ac & 16 & 16 & 1263.0 & 1310.7 & 1288.8 & 1157.0 & 0.97 & 0.95 & 0.93 & 0.90 \\        
        \bottomrule
    \end{tabular}
    \caption{Comparison between pure random and IEEE 802.11ac channel allocation schemes. The value of $W$ is selected uniformly at random between $20$ and $W_{\max}$.}\label{Tbl:11acChannels}
\end{table*}

\subsection{Final Remarks}

Figure~\ref{Fig:ResultsInWLANs} shows that a random channel selection greatly affects the overall system performance. For instance, with $10$ WLANs, the expected throughput achieved by a single WLAN using the waterfilling algorithm is almost $100$ \% higher than the expected throughput achieved with $W_{\max}=40$ MHz, which is the channel width that gives the best throughput when the channel position is selected randomly. Similarly, the JFI value is significantly lower than that obtained by the waterfilling algorithm. The spectrum utilisation is also lower  because several WLANs use the same basic channels, while others remain empty. Similar observations can be made from Figure \ref{Fig:ResultsIncChannels}.

Therefore, there is an important gap between the performance of the centralized channel allocation algorithm and the performance when each WLAN randomly selects the channel to use. There are several possible solutions to reduce this gap for autonomous WLANs: 1) Use a database in the cloud to store information about the channels that are used in each geographical area. This database could be used to find empty channels for new WLANs. However, there is no way to force already existing WLANs to adapt to an increasing demand in that area and reduce their channel width. 2) Use a decentralised channel selection algorithm that is able to adapt to the spectrum occupancy based on the instantaneous information that it is able to infer from the behaviour of the neighboring WLANs. 


\section{Conclusions} \label{Sec:Conclusions}

In this paper, we have introduced, described and characterised the interactions that occur in the operation of multiple neighboring WLANs when they use channel bonding. To capture these interactions, we have developed and validated an analytical framework based on a CTMN model. This framework was then used to evaluate the system performance in terms of the number of neighboring WLANs, the number of basic channels available and the set of channel widths that each WLAN is allowed to use. We have also proposed a centralised waterfilling algorithm that provides a proportional fair global channel allocation, or at least the best suboptimal allocation, because we are dealing with a discrete state space. 

The results obtained when WLANs select the channel center frequency and the channel width randomly, which is a good representation of what occurs in real deployments, show that the throughput, the spectrum utilisation and the fairness are significantly lower than the values obtained using the centralised algorithm. This indicates the need to develop smarter decentralised channel selection algorithms to make an efficient use of the available spectrum for autonomous WLANs.


\bibliographystyle{IEEEtran}
\bibliography{TheBib}

\begin{thebibliography}{10}
\providecommand{\url}[1]{#1}
\csname url@samestyle\endcsname
\providecommand{\newblock}{\relax}
\providecommand{\bibinfo}[2]{#2}
\providecommand{\BIBentrySTDinterwordspacing}{\spaceskip=0pt\relax}
\providecommand{\BIBentryALTinterwordstretchfactor}{4}
\providecommand{\BIBentryALTinterwordspacing}{\spaceskip=\fontdimen2\font plus
\BIBentryALTinterwordstretchfactor\fontdimen3\font minus
  \fontdimen4\font\relax}
\providecommand{\BIBforeignlanguage}[2]{{%
\expandafter\ifx\csname l@#1\endcsname\relax
\typeout{** WARNING: IEEEtran.bst: No hyphenation pattern has been}%
\typeout{** loaded for the language `#1'. Using the pattern for}%
\typeout{** the default language instead.}%
\else
\language=\csname l@#1\endcsname
\fi
#2}}
\providecommand{\BIBdecl}{\relax}
\BIBdecl

\bibitem{deek2011impact}
L.~Deek, E.~Garcia-Villegas, E.~Belding, S.-J. Lee, and K.~Almeroth, ``{The
  Impact of Channel Bonding on 802.11n Network Management},'' in
  \emph{Proceedings of the Seventh COnference on emerging Networking
  EXperiments and Technologies (CONEXT)}.\hskip 1em plus 0.5em minus
  0.4em\relax ACM, 2011.

\bibitem{IEEE80211n}
``{IEEE 802.11n. Standard for Wireless LAN Medium Access Control (MAC) and
  Physical Layer (PHY): Enhancements for High Throughput},'' \emph{IEEE}, 2009.

\bibitem{IEEE80211ac}
I.~P802.11ac, ``{Standard for Wireless LAN Medium Access Control (MAC) and
  Physical Layer (PHY) specifications: Enhancements for Very High Throughput
  for Operation in Bands below 6 GHz.}'' \emph{IEEE}, 2014.

\bibitem{bellalta2015ieee}
B.~Bellalta, ``{IEEE 820.11ax: High-Efficiency WLANs},'' \emph{arXiv preprint
  arXiv:1501.01496}, 2015.

\bibitem{boorstyn1987throughput}
R.~Boorstyn, A.~Kershenbaum, B.~Maglaris, and V.~Sahin, ``{Throughput Analysis
  in Multihop CSMA Packet Radio Networks},'' \emph{IEEE Transactions on
  Communications}, vol.~35, no.~3, pp. 267--274, 1987.

\bibitem{bianchi2000performance}
G.~Bianchi, ``{Performance Analysis of the IEEE 802.11 Distributed Coordination
  Function},'' \emph{IEEE Journal on Selected Areas in Communications},
  vol.~18, no.~3, pp. 535--547, 2000.

\bibitem{arslan2010auto}
M.~Y. Arslan, K.~Pelechrinis, I.~Broustis, S.~V. Krishnamurthy, S.~Addepalli,
  and K.~Papagiannaki, ``{Auto-configuration of 802.11n WLANs},'' in
  \emph{Proceedings of the 6th International COnference on emerging Networking
  EXperiments and Technologies (CONEXT)}.\hskip 1em plus 0.5em minus
  0.4em\relax ACM, 2010.

\bibitem{herzen2013distributed}
J.~Herzen, R.~Merz, and P.~Thiran, ``{Distributed Spectrum Assignment for Home
  WLANs},'' in \emph{IEEE INFOCOM}.\hskip 1em plus 0.5em minus 0.4em\relax
  IEEE, 2013, pp. 1573--1581.

\bibitem{Gong2011ChannelBounding}
M.~X. Gong, B.~Hart, L.~Xia, and R.~Want, ``{Channel Bounding and MAC
  Protection Mechanisms for 802.11ac.}'' in \emph{IEEE GLOBECOM}.\hskip 1em
  plus 0.5em minus 0.4em\relax IEEE, 2011, pp. 1--5.

\bibitem{park2011ieee}
M.~Park, ``{IEEE 802.11ac: Dynamic Bandwidth Channel Access},'' in \emph{IEEE
  ICC}.\hskip 1em plus 0.5em minus 0.4em\relax IEEE, 2011, pp. 1--5.

\bibitem{bellaltachannel}
B.~Bellalta, A.~Faridi, J.~Barcelo, A.~Checco, and P.~Chatzimisios, ``{Channel
  Bonding in Short-Range WLANs},'' in \emph{Proceedings of 20th European
  Wireless Conference European Wireless}.\hskip 1em plus 0.5em minus
  0.4em\relax IEEE, 2014, pp. 1--7.

\bibitem{leung2003frequency}
K.~Leung and B.~Kim, ``{Frequency Assignment for IEEE 802.11 Wireless
  Networks},'' in \emph{{{IEEE 58th Vehicular Technology Conference
  (VTC)}}}.\hskip 1em plus 0.5em minus 0.4em\relax IEEE, 2003.

\bibitem{mishra2006client}
A.~Mishra, V.~Brik, S.~Banerjee, A.~Srinivasan, and W.~Arbaugh, ``{{A
  Client-driven Approach for Channel Management in Wireless LANs}},'' in
  \emph{{{IEEE INFOCOM}}}.\hskip 1em plus 0.5em minus 0.4em\relax IEEE, 2006,
  pp. 1--12.

\bibitem{mishra2006distributed}
A.~Mishra, V.~Shrivastava, D.~Agrawal, S.~Banerjee, and S.~Ganguly,
  ``{Distributed Channel Management in Uncoordinated Wireless Environments},''
  in \emph{Proceedings of the 12th Annual International Conference on Mobile
  Computing and Networking}, ser. MobiCom '06.\hskip 1em plus 0.5em minus
  0.4em\relax New York, NY, USA: ACM, 2006, pp. 170--181.

\bibitem{raniwala2005architecture}
A.~Raniwala and T.~Chiueh, ``{{Architecture and Algorithms for an IEEE
  802.11-based Multi-channel Wireless Mesh Network}},'' in \emph{{{IEEE
  INFOCOM}}}, vol.~3, 2005, pp. 2223 -- 2234.

\bibitem{clifford2007channel}
P.~Clifford and D.~J. Leith, ``{Channel dependent interference and
  decentralized colouring},'' in \emph{Network Control and Optimization}.\hskip
  1em plus 0.5em minus 0.4em\relax Springer, 2007, pp. 95--104.

\bibitem{kauffmann2007measurement}
B.~Kauffmann, F.~Baccelli, A.~Chaintreau, V.~Mhatre, K.~Papagiannaki, and
  C.~Diot, ``{{Measurement-based Self Organization of Interfering 802.11
  Wireless Access Networks}},'' in \emph{{{IEEE INFOCOM}}}.\hskip 1em plus
  0.5em minus 0.4em\relax IEEE, 2007, pp. 1451--1459.

\bibitem{kauffmann2007self}
------, ``{Self Organization of Interfering 802.11 Wireless Access Networks},''
  [Research Report] RR-5649, 2005, pp.25. INRIA, year = {2007}, Tech. Rep.

\bibitem{durvy2006packing}
M.~Durvy and P.~Thiran, ``A packing approach to compare slotted and non-slotted
  medium access control.'' in \emph{IEEE INFOCOM}.\hskip 1em plus 0.5em minus
  0.4em\relax IEEE, 2006, pp. 1--12.

\bibitem{liew2010back}
S.~C. Liew, C.~H. Kai, H.~C. Leung, and P.~Wong, ``{Back-of-the-envelope
  Computation of Throughput Distributions in CSMA Wireless Networks},''
  \emph{IEEE Transactions on Mobile Computing}, vol.~9, no.~9, pp. 1319--1331,
  2010.

\bibitem{nardelli2012closed}
B.~Nardelli and E.~W. Knightly, ``{Closed-form throughput expressions for CSMA
  networks with collisions and hidden terminals},'' in \emph{IEEE
  INFOCOM}.\hskip 1em plus 0.5em minus 0.4em\relax IEEE, 2012, pp. 2309--2317.

\bibitem{laufercapacity}
R.~Laufer and L.~Kleinrock, ``On the capacity of wireless csma/ca multihop
  networks,'' in \emph{INFOCOM}.\hskip 1em plus 0.5em minus 0.4em\relax IEEE,
  2013, pp. 1312 -- 1320.

\bibitem{wang2005throughput}
X.~Wang and K.~Kar, ``{Throughput Modelling and Fairness Issues in CSMA/CA
  based Ad-hoc Networks},'' in \emph{IEEE INFOCOM}, vol.~1.\hskip 1em plus
  0.5em minus 0.4em\relax IEEE, 2005, pp. 23--34.

\bibitem{bellalta2014bookchapter}
B.~Bellalta, A.~Zocca, C.~Cano, A.~Checco, J.~Barcelo, and A.~Vinel,
  ``{Throughput Analysis in CSMA/CA Networks Using Continuous Time Markov
  Networks: A Tutorial},'' \emph{Wireless Networking for Moving Objects:
  Protocols, Architectures, Tools, Services and Applications. Springer}, vol.
  8611, p. 115, 2014.

\bibitem{bellaltaperformance}
B.~Bellalta, E.~Belyaev, M.~Jonsson, and A.~Vinel, ``{Performance Evaluation of
  IEEE 802.11p-Enabled Vehicular Video Surveillance System},'' \emph{IEEE
  Communication Letters}, vol.~18, pp. 708--711, 2014.

\bibitem{cali2000ieee}
F.~Cali, M.~Conti, and E.~Gregori, ``{IEEE 802.11 Protocol: Design and
  Performance Evaluation of an Adaptive Backoff Mechanism},'' \emph{IEEE
  Journal on Selected Areas in Communications}, vol.~18, no.~9, pp. 1774--1786,
  2000.

\bibitem{deng2004tuning}
J.~Deng, B.~Liang, and P.~K. Varshney, ``{Tuning the carrier sensing range of
  IEEE 802.11 MAC},'' in \emph{IEEE GLOBECOM}, vol.~5.\hskip 1em plus 0.5em
  minus 0.4em\relax IEEE, 2004, pp. 2987--2991.

\bibitem{cai2013improving}
Y.~Cai, K.~Xu, Y.~Mo, B.~Wang, and M.~Zhou, ``{Improving WLAN Throughput via
  Reactive Jamming in the Presence of Hidden Terminals},'' in \emph{IEEE
  Wireless Communications and Networking Conference (WCNC)}.\hskip 1em plus
  0.5em minus 0.4em\relax IEEE, 2013, pp. 1085--1090.

\bibitem{jang2012ieee}
B.~Jang and M.~L. Sichitiu, ``{IEEE 802.11 Saturation Throughput Analysis in
  the Presence of Hidden Terminals},'' \emph{IEEE/ACM Transactions on
  Networking (TON)}, vol.~20, no.~2, pp. 557--570, 2012.

\bibitem{chatzimisios2004performance}
P.~Chatzimisios, A.~C. Boucouvalas, and V.~Vitsas, ``Performance analysis of
  ieee 802.11 dcf in presence of transmission errors,'' in \emph{IEEE ICC},
  vol.~7.\hskip 1em plus 0.5em minus 0.4em\relax IEEE, 2004, pp. 3854--3858.

\bibitem{kelly1979reversibility}
F.~P. Kelly, \emph{{Reversibility and Stochastic Networks}}.\hskip 1em plus
  0.5em minus 0.4em\relax Wiley, New York, 1979.

\bibitem{garetto2006modeling}
M.~Garetto, T.~Salonidis, and E.~W. Knightly, ``{Modeling Per-flow Throughput
  and Capturing Starvation in CSMA Multi-hop Wireless Networks},''
  \emph{IEEE/ACM Trans. Netw.}, vol.~16, no.~4, pp. 864--877, Aug. 2008.

\bibitem{van2010insensitivity}
P.~M. van~de Ven, S.~C. Borst, J.~Van~Leeuwaarden, and A.~Prouti{\`e}re,
  ``Insensitivity and stability of random-access networks,'' \emph{Performance
  Evaluation}, vol.~67, no.~11, pp. 1230--1242, 2010.

\bibitem{heusse2003performance}
M.~Heusse, F.~Rousseau, G.~Berger-Sabbatel, and A.~Duda, ``{Performance Anomaly
  of 802.11b},'' in \emph{IEEE INFOCOM}, vol.~2.\hskip 1em plus 0.5em minus
  0.4em\relax IEEE, 2003, pp. 836--843.

\bibitem{radunovic2002unified}
B.~Radunovi\'{c} and J.-Y.~L. Boudec, ``{A Unified Framework for Max-min and
  Min-max Fairness with Applications},'' \emph{IEEE/ACM Trans. Netw.}, vol.~15,
  no.~5, pp. 1073--1083, Oct. 2007.

\bibitem{chen2002reusing}
G.~Chen and B.~K. Szymanski, ``{Reusing Simulation Components: COST: A
  Component-oriented Discrete Event Simulator},'' in \emph{Proceedings of the
  34th Conference on Winter Simulation: Exploring New Frontiers}, ser. WSC
  '02.\hskip 1em plus 0.5em minus 0.4em\relax Winter Simulation Conference,
  2002, pp. 776--782.

\end{thebibliography}


\begin{appendices}

\section{System Parameters and Scenario Considerations}\label{Sec:Parameters}

We assume that all WLANs operate using the IEEE 802.11ac amendment \cite{IEEE80211ac}. Therefore, the WLANs operate in the $5$ GHz ISM band, where each basic channel has a width of $20$ MHz. $W$ can take values from  $\{20,~40,~80,~160\}$ MHz. In other words, the channel $C_i$, selected by WLAN $i$ can be composed of $c_i \in \{ 1, 2, 4, 8\}$ basic channels. The parameters considered are shown in Table \ref{Tbl:parameters}. Moreover, we consider that for each value of $c$, WLANs use a different modulation and coding rate, as shown in Table \ref{Tbl:parameters_channels}. Unless otherwise stated, all WLANs have two nodes.

\begin{table}[thhhhhhhhhhhhhhh!!]
\centering
 \begin{tabular}{lcc}
   \toprule
  {\bf Parameter} & {\bf Notation} & {\bf Value} \\
  \midrule
  Packet Length & $L_d$ & $12000$ bits \\ 
  Number of aggregated packets & $N_{\text{a}}$ & $64$ packets \\ 
  Number of SU-MIMO spatial streams & $N_{\text{su}}$ & $2$ packets \\ 
  Backoff Contention Window & CW & $16$ slots \\ 
  Slot Duration & $T_{\text{slot}}$ & $9~\mu$s \\
  Average time decreasing the backoff & $E[B]$ & $\frac{\text{CW}}{2}\cdot T_{\text{slot}}$\\ 
  DIFS & - & $34~\mu$s\\
  \bottomrule
 \end{tabular}
 \caption{{Parameter values based on IEEE 802.11ac}}\label{Tbl:parameters}
\end{table}

\begin{table*}[thhh!!]
\centering    
 \begin{tabular}{lccccc}
   \toprule
  {\bf $c$} & {\bf Data Subcarriers, $\xi(c)$} & {\bf Modulation, $N_{\text{m}}$} & {\bf Coding Rate, $N_{\text{c}}$} & {\bf Tx Rate (Mbps)}  & {\bf Min. Sensitivity (dBm)}  \\
  \midrule
  1 & $52$ & 64-QAM, $6$ bits & $5/6$ & $65$ & $-64$ \\
  2 & $108$ & 64-QAM, $6$ bits & $3/4$ & $121.5$ & $-62$\\
  4 & $234$ & 16-QAM, $4$ bits & $3/4$ & $175.5$ & $-61$\\
  8 & $468$ & 16-QAM, $4$ bits & $1/2$ & $232$ & $-65$\\  
  \bottomrule
 \end{tabular}
 \caption{Transmission Rates for each number of basic channels. This values are computed for a packet error probability less than 10 \% for 4096 Bytes packets \cite{IEEE80211ac}}\label{Tbl:parameters_channels}
\end{table*}

All nodes are equipped with at least two antennas, which they use to transmit two spatial streams in single-user MIMO mode. Packet aggregation is also considered, and $64$ packets are included in each transmission. Under these conditions, the time required to transmit a packet by node $j$ in WLAN $i$ is $T_{i,j}(c_i,\gamma_{i,j},L_{i,j})$, which is computed as follows 
\begin{align}\label{Eq:Ts}
	& T_{i,j}(c_i,\gamma_{i,j},L_{i,j})=  \nonumber \\& =  \left(T_{\text{PHY}} + \left\lceil \frac{\text{SF} + N_{\text{a}} (\text{MD}+\text{MH}+L_{i,j}) + \text{TB}}{N_{\text{su}}L_{\text{DBPS}}(c_i,\gamma_{i,j})}\right\rceil T_s \right) \nonumber +\\ &+\text{SIFS}+\left(T_{\text{PHY}}+\left \lceil \frac{\text{SF} + L_{\text{BA}} + \text{TB} }{L_{\text{DBPS}}(1,\gamma)} \right \rceil T_s \right)+\text{DIFS} + T_{\text{slot}},
\end{align}
where $T_{\text{PHY}}=40 \mu$s is the duration of the PHY-layer preamble and headers, $T_s=4~\mu$s is the duration of an OFDM (Orthogonal Frequency Division Multiplexing) symbol. {SF is the \textit{service field} ($16$ bits), $\text{MD}$ is the \textit{MPDU Delimiter} ($32$ bits) MH is the \textit{MAC header} ($288$ bits), TB is the number of \textit{tail bits} ($6$ bits), and $L_{\text{BA}}$ is the \textit{Block-ACK} length ($256$ bits). $L_{\text{DBPS}}(c,\gamma)=N_\text{m}(\gamma) N_\text{c}(\gamma) \xi(c)$ is the number of bits in each OFDM symbol, where $N_\text{m}(\gamma)$ is the number of bits per modulation symbol, $N_\text{c}(\gamma)$ is the coding rate, and $\xi(c)$ is the number of data subcarriers when $c$ basic channels are bonded together}. $N_{\text{su}}=2$ is the number of single-user MIMO streams, and $N_{\text{a}}=64$ is the number of packets that are aggregated in each transmission.  

Lastly, we consider that in all WLANs the STAs are located near the AP. In that situation, the packet error probability is assumed to be negligible. 


In all the plots, unless otherwise stated, each point is the average result of $2000$ different randomly generated scenarios, where each scenario represents a single generated global channel allocation $\mathcal{C}$. This number of realisations guarantees that the standard deviation of the error in the sample mean relative to the true mean of the computed throughput is less than $10$ Mbps, which is considered an acceptable error. 

\subsection{Simulation Tool}

To validate the analytical model, a simulator of the described scenario was built based on the COST (Component Oriented Simulation Toolkit) libraries \cite{chen2002reusing}. The simulator accurately reproduces the described scenario and the operation of each node, including the slotted backoff mechanism that is considered in IEEE 802.11 WLANs, the presence of collisions and the capture effect when multiple packets are simultaneously received with very different power levels \cite{deng2004tuning}. In the simulator, when the capture effect is enabled, we assume that collisions between packets from nodes that belong to different WLANs do not cause the loss of the transmitted packets at the corresponding receiver.  

By comparing the results obtained from the simulator with those obtained from the analytical model, we can assess the model's accuracy and the impact of the assumptions that were required to construct it.

\end{appendices}

\end{document}